\documentclass[aps,prb,twocolumn,showpacs,final,floatfix]{revtex4-1}
\usepackage{amsmath}
\usepackage{amsfonts}
\usepackage{amssymb}
\usepackage{physics}
\usepackage{color}
\usepackage{hyperref}
\usepackage{inputenc}
\usepackage{enumerate}
\usepackage{graphicx}
\usepackage{float}
\usepackage{amsmath}
\usepackage{amsfonts}
\usepackage{lipsum}
\usepackage{dcolumn}
\usepackage{hyperref}
\usepackage{subfigure}
\usepackage{epsfig}
\usepackage{epsf} 
\usepackage{epstopdf}

\usepackage{color}
\usepackage{enumitem}
\newcommand\la{\langle}
\newcommand\ra{\rangle}
\newcommand\cH{\mathcal{H}}
\begin{document}


\title{Transport in spinless superconducting wires}


\author{Junaid Majeed Bhat}

\author{Abhishek Dhar}
\affiliation{International Centre for Theoretical Sciences,  Tata Institute of Fundamental Research, Bengaluru-560089, India}


\date{\today}

\begin{abstract}
We consider electron transport in a model of a spinless superconductor described by a Kitaev type lattice Hamiltonian where the electron interactions are modelled through a superconducting  pairing term. The superconductor is sandwiched between two normal metals kept at different temperatures and chemical potentials and  are themselves modelled as non-interacting spinless fermions. For this set-up we compute the exact steady state properties of the system using the quantum Langevin equation approach. Closed form exact expressions for particle current, energy current and other two-point correlations are obtained in the Landauer-type forms and involve two nonequilibrium Green's functions.  The current expressions are found out to be sum of three terms having simple physical interpretations.  We then discuss a numerical approach where we construct the time-evolution of the two point correlators of the system from the eigenspectrum of the complete quadratic Hamiltonian describing the system and leads. By starting from an initial state corresponding to the leads in thermal equilibrium and the system in an arbitrary state, the long time solution for the correlations, before recurrence times, gives us steady state  properties. We use this independent numerical method for verifying the results of the exact solution. We also investigate analytically the presence of high energy bound states and obtain expressions for their  contributions to two point correlators. As  applications of our general formalism we present results on thermal conductance and on the conductance of a Kitaev chain with next nearest neighbour interactions which allows topological phases with different winding numbers.

\end{abstract}

\maketitle

\section{Introduction}
\label{sec:intro}
The Kitaev chain  models a one-dimensional spinless p-wave superconductor~\cite{kitaev} and provides one of the simplest examples of a topological insulator. This system has the  so-called Majorana Bound States (MBS), which are topologically protected   zero-energy bound states, localised at the boundaries of an open chain. Several proposals were put forward~\cite{fu2008superconducting,sau2010generic,lutchyn2010majorana,oreg2010helical,sau2010non} to realise  the Kitaev chain experimentally and observe the MBS.  Some of these experimental proposals have already been successfully implemented~\cite{mourik2012ep,das2012zero,deng2012anomalous,rokhinson2012fractional,finck2013anomalous,churchill2013superconductor}. 
 One of the key experimental  signatures of the MBS is the zero-bias peak in the differential tunnelling conductance   and  Ref.~[\onlinecite{mourik2012ep,das2012zero,deng2012anomalous}] were some of first experiments which reported evidence for this peak. 
However, these experiments show several  deviations from the expected theoretical results. For example, the strength of the zero-bias peak was found  to be much smaller than the expected value $2e^2/h$. A comprehensive discussion of some of the issues can be found in Ref.~[\onlinecite{lobos2015,aguado2017majorana}]. More recent experiments with  barrier free contacts between the normal metal and the superconductor, along with  implementation of  better fabrication techniques have helped in resolving some of these deviations~\cite{aguado2017majorana}. More interesting topological phases are revealed as one goes beyond the nearest neighbour Hamiltonian in a Kitaev chain~\cite{niu2012majorana, uhrich2020out,vodola2015long}. In particular Ref.~\onlinecite{niu2012majorana} shows the existence of two different topologically non-trivial phases in a 1-D Kitaev chain with next to nearest neighbour couplings. Such models are of interest since  interacting spin systems with external fields can be mapped to these using a Jordan Wigner transformation. Another such example can be found in Ref.~\onlinecite{nasu2017thermal} where a spin system on a Honeycomb lattice is mapped to a Kitaev type lattice Hamiltonian. This study shows interesting results for thermal conductance in the linear response regime. Therefore, a complete understanding of the transport properties of such  superconducting wires defined on a general lattice  is very important for our understanding of the transport characteristics of the different topological phases. 

We note that  many of the interesting topological characteristics can be understood in the context of the isolated system, in either the geometry with free boundary conditions or one with periodic boundary conditions. One can also extract interesting results on transport within the framework of  the Green-Kubo formalism~\cite{shankar2018}. However experiments on transport are typically done with leads and reservoirs, kept at different chemical potentials and temperatures, and understanding this requires the framework of open quantum systems. One could expect  that the MBS at the ends of a Kitaev chain would get affected on coupling the wire to infinite leads (made of normal material). 
Some of the open system approaches that have been used to understand transport in superconducting wires include  scattering approach~\cite{btk,Sengupta2001,diptiman2015,maiellaro2019}, nonequilibrium Keldysh Green's function (NEGF) approach~\cite{lobos2015,doornenbal2015conductance,komnik2016} and the quantum Langevin equation approach~\cite{roy2012}. 
One of the earliest  study of  electron transport through superconducting channels was by  Blonder, Tinkham, and Klapwijk~\cite{btk}. In their approach, the electron transport properties of a superconducting channel were studied by considering scattering of a plane wave at a junction separating normal metal and the superconductor(NS junction) and the important role of Andreev reflection was pointed out. A similar approach has been used in Ref.~[\onlinecite{diptiman2015}] to study  conductance of a one-dimensional system consisting of a p-wave superconductor connected to leads at the two ends (NSN junction).  A very recent study on thermal transport of 1-D Kitaev chain shows a zero bias peak in thermal conductance very close to the phase transition point \cite{bondyopadhaya2020nonequilibrium}. In the present work we present very general and compact formulas for both particle and thermal conductance. We show that this peak arises  due to the divergence of the localization length of the  Majorana zero modes as one approaches the transition point, and we discuss related finite-size effects.

Another numerical approach that has been used to study transport in superconducting wires connected to normal wires uses the fact that the models involve quadratic Hamiltonians and hence a direct diagonalization is possible for quite long chains. In a recent work~\cite{roy2019}, this approach was used to study transport in various configurations of normal and superconducting wires  and it was noted that bound states could lead to persistent oscillating currents in the system.  We investigate this issue analytically and present expressions for the bound state contributions from which the oscillation frequencies can be  read off directly. Note that these bound states are different from the topological MBS and are found as a discrete energy level outside the continuous band of the spectrum of the entire system(wire+baths). The MBS on the other hand lie within the continuous band.

The quantum Langevin equations (QLE) approach~\cite{ford1965statistical} provides one of the most direct and intuitive approaches for open quantum systems. Here one writes the effective dynamics of a system coupled to thermal and particle reservoirs, and this can be  used to  obtain the NEGF results, for both electronic and phononic systems~\cite{dhar2006,dhar2006heat}.
The nonequilibrium particle transport properties of  the Kitaev chain  were first studied using  the QLE-NEGF formalism in Ref.~[\onlinecite{roy2012}] and several interesting results were obtained. In particular, the  variation of differential conductance  with the  wire length and the current voltage characteristics of the wire were studied.

 In the  present work  we extend the analysis in  Ref.~[\onlinecite{roy2012}] to obtain  the nonequilibrium steady state (NESS) properties of a spinless superconductor defined on  an arbitrary lattice in any dimension, the Kitaev chain then occurs as a special case. Using the QLE-NEGF formalism, we obtain exact formal  expressions for the particle current, heat current, density and other two-point correlations in terms of  nonequilibrium Green's functions. We  also obtain analytically the bound state corrections to these two point correlations. We find that in the steady state all quantities can be expressed in terms of a pair of retarded and advanced Green's functions $G_1^\pm(\omega)$ and $G_2^\pm(\omega)$.   As a result of this the currents are obtained as a sum of three different terms each of which has a simple and definite physical interpretation. Secondly, noting that the system is described by a quadratic Hamiltonian, we use an exact numerical scheme to study the microscopic time evolution of the coupled system and reservoirs, starting from a product initial condition. Using this numerical scheme for the Kitaev chain, we verify the analytic results from the QLE-NEGF formalism, for chains of finite lengths as well as the expression for the bound state contributions.  We comment on various  issues that arise in parameter regimes where the combined system-reservoirs has  high energy bound states (distinct from the MBS). Finally  as an application of our  general formalism, we discuss the transport properties of an open Kitaev chain with  interaction couplings that extend beyond nearest neighbours.

This paper is structured as follows. In Sec.~\eqref{sec:QLE}, we define the precise model of a spinless superconductor (which we refer to as the Kitaev wire) coupled to two thermal reservoirs made of free fermions. Starting from  the Heisenberg equations of motion, we obtain the effective quantum Langevin equations for the wire.  In Sec.~\eqref{sec:ness}, these Langevin equations of motion are solved using Fourier transform to get the steady state solution for the current and two-point correlations in the wire. In Sec.~\eqref{sec:timeevol}, we present a different  formal numerical scheme to compute two-point correlators at all times using the exact diagonalization of  the quadratic Hamiltonian describing the superconductor and the high energy bound state contributions to the two point correlators discussed in Sec.~\ref{sec:ness}. With the numerical scheme we obtain the current and the densities at any finite time. For finite systems, this method is numerically implemented in Sec.~\eqref{sec:numerics}, for the special case of a one-dimensional Kitaev chain (connected to normal baths) and a comparison is made with the steady state results obtained using the methods of Sec.~\eqref{sec:QLE}.   The results on the  conductance of the Kitaev chain with  next nearest neighbour interactions and multiple topological phases is discussed in Sec.~\eqref{sec:long}. We conclude with a discussion in Sec.~\eqref{sec:summary}.

\section{Quantum Langevin equations and Green's function formalism}
\label{sec:QLE}
We consider a wire coupled to  thermal baths on its  two ends. The wire Hamiltonian, $\mathcal{H}^W$, is taken to correspond to a spinless  superconductor  while the two baths  are modelled by the tight binding Hamiltonians, $\mathcal{H}^L$ and $\mathcal{H}^R$. The $L$ and $R$  superscripts denote the baths on the left and the right of the wire respectively. The couplings of the wire with the two baths, $\mathcal{H}^{WR}$ and $\mathcal{H}^{WL}$, are also modelled by tight binding Hamiltonians.  Let us denote by $\lbrace c_m, c_m^\dagger\rbrace$, $\lbrace c_\nu, c_\nu^\dagger\rbrace$ and $\lbrace c_{\nu^\prime}, c_{\nu^\prime}^\dagger\rbrace$ the  annihilation and creation operators of the system, left bath and the right bath respectively. These satisfy usual fermionic anti-commutation relations. For lattice sites on the bath we use the Latin indices, ${i,j,...}$, for sites on the left reservoir we use the Greek indices, $\alpha, \nu,...$,  and for sites on the right reservoir, the primed Greek letters $\alpha', \nu',...$. We take the Hamiltonian of the full system of  wire and baths  as follows:
\begin{equation}
\mathcal{H}=\mathcal{H}^W+\mathcal{H}^{WL}+\mathcal{H}^{WR}+\mathcal{H}^{L}+\mathcal{H}^{R}, \label{hamil}
\end{equation}
where
\begin{align}
\mathcal{H}^W&=\sum_{mn} H^W_{mn}c_m^\dagger c_n+\Delta_{mn}c_m^\dagger c_n^\dagger+\Delta_{mn}^\dagger c_mc_n,  \\
\mathcal{H}^{WL}&=\sum_{\nu m}V^L_{m\nu}c_m^\dagger   c_\nu+V^{L^\dagger}_{\nu m} c_\nu^\dagger c_m,\\
\mathcal{H}^{WR}&=\sum_{\nu^\prime m}V^R_{m\nu^\prime}c_m^\dagger   c_{\nu^\prime}+V^{R^\dagger}_{\nu^\prime m} c_{\nu^\prime}^\dagger c_m,\\
\mathcal{H}^L&=\sum_{\mu\nu} H^L_{\mu\nu}c_\mu^\dagger c_\nu\\\mathcal{H}^R&=\sum_{\mu^\prime\nu^\prime} H^R_{\mu^\prime\nu^\prime}c_{\mu^\prime}^\dagger c_{\nu^\prime}.
\end{align} 
 The model considered here is quite general in the sense that we allow non-zero hopping elements between arbitrary sites and similarly the superconducting pairing term is allowed between any pair of sites. 
Thus there are no restrictions on  dimensionality and the  structure of the underlying lattice and the range of the interactions. The results for the one-dimensional Kitaev chain with nearest neighbour interactions  follows as a special case.

We now follow the approach of  Ref.~[\onlinecite{dhar2006}] to obtain the NEGF-type results for this system. First we note that the Heisenberg equations of motion for the wire sites and bath sites are given by:
\begin{align}
\notag\dot{c}_l&=-i\sum_m H^W_{lm}c_m-i\sum_mK_{lm}c_m^\dagger\\&\hspace{2cm}-i\sum_{\alpha}V^L_{l\alpha }c_\alpha-i\sum_{\alpha^\prime}V^R_{l\alpha^\prime }c_{\alpha^\prime},\label{hes}\\
\dot{c}_{\alpha}&=-i\sum_{\nu}H^L_{\alpha\nu}c_{\nu}-i\sum_l V^{L\dagger}_{\alpha l}c_l\label{hebl},\\
\dot{c}_{\alpha^\prime}&=-i\sum_{\nu^\prime}H^R_{\alpha^\prime\nu^\prime}c_{\nu}-i\sum_l V^{R\dagger}_{\alpha^\prime l}c_l,\label{hebr}
\end{align}
where $K_{lm}=(\Delta-\Delta^T)_{lm}$. We treat the term containing $c_l$ in Eq.~\eqref{hebl} and Eq.~\eqref{hebr} as the inhomogeneous parts and solve these equations  using the following Green's functions corresponding to the homogeneous part of the equations:
\begin{align}
g_L^+(t)&=-ie^{-itH^L}\theta(t)=\int_{-\infty}^\infty \frac{d\omega}{2\pi}g_L^+(\omega)e^{-i\omega t},\\
g_R^+(t)&=-ie^{-itH^R}\theta(t)=\int_{-\infty}^\infty \frac{d\omega}{2\pi}g_R^+(\omega)e^{-i\omega t}.
\end{align}
In terms of these, we obtain the following solutions for the reservoir equations (for $t>t_0$):
\begin{align}
c_{\alpha}(t)&=i\sum_{\nu}[g_L^+(t-t_0)]_{\alpha\nu}c_\nu(t_0)\notag\\&\hspace{2cm}+\int_{t_0}^{t}ds \sum_{\nu l}[g_L^+(t-s)]_{\alpha \nu} V^{L^\dagger}_{\nu l} c_l(s),\\
c_{\alpha^\prime}(t)&=i\sum_{\nu^\prime}[g_R^+(t-t_0)]_{\alpha^\prime\nu^\prime}c_{\nu^\prime}(t_0)\notag\\&\hspace{2cm}+\int_{t_0}^{t}ds \sum_{\nu^\prime l}[g_R^+(t-s)]_{\alpha^\prime\nu^\prime} V^{R^\dagger}_{\nu^\prime l} c_l(s).
\end{align}
Substituting these results in the Heisenberg equation for the wire sites we have:
\begin{align}
\notag\dot{c}_l&=-i\sum_m H^W_{lm}c_m-i\sum_mK_{lm}c_m^\dagger-i\eta^L_l-i\eta_l^R\\&\hspace{1cm}-i\int_{t_0}^{t}ds\sum_{\alpha\nu m}V^L_{l\alpha}[g_L^+(t-s)]_{\alpha \nu}V^{L^\dagger}_{\nu m} c_m(s)\notag\\&\hspace{1cm}-i\int_{t_0}^{t}ds\sum_{\alpha^\prime\nu^\prime m}V^R_{l\alpha^\prime}[g_R^+(t-s)]_{\alpha^\prime \nu^\prime}V^{R^\dagger}_{\nu^\prime m} c_m(s),\label{les}
\end{align}
where
\begin{align}
\eta_l^L&=i\sum_{\alpha\nu}V^L_{l\alpha}[g_L^+(t-t_0)]_{\alpha\nu}c_\nu(t_0),\\
\eta_l^R&=i\sum_{\alpha^\prime\nu^\prime}V^R_{l\alpha^\prime}[g_L^+(t-t_0)]_{\alpha^\prime\nu^\prime}c_{\nu^\prime}(t_0).
\end{align}
At $t=t_0$, we choose the two reservoirs to be described by  grand canonical ensembles at temperatures and chemical potentials given by $(T_L,\mu_L)$ and $(T_R,\mu_R)$ respectively. This allows us to determine the correlation  properties of the terms $\eta_l^L$ and $\eta_l^R$. For the left bath we have:
\begin{align}
\expval{\eta_l^L(t)\eta_m^L(t^\prime)}&=\expval{\eta_l^{L\dagger}(t)\eta_m^{L\dagger}(t^\prime)}=0,\\
\notag\expval{\eta_l^{L\dagger}(t)\eta_m^L(t^\prime)}&=\sum_{\alpha\mu\nu\sigma}V^{L*}_{l\alpha}[g^{+*}_L(t-t_0)]_{\alpha\nu}V^L_{m\mu}\\&\hspace{1cm}[g^{+}_L(t^\prime-t_0)]_{\mu\sigma}\expval{c_\nu^\dagger(t_0)c_\sigma(t_0)},\label{ncorr}
\end{align} 
with similar expressions  for $\eta_l^R$. We thus see that Eq.~\eqref{les} has the structure of a  quantum Langevin equation for the wire where the reservoir contributions are split into  noise  (terms given by $\eta_l^L$ and $\eta_l^R$), and dissipation (the terms in Eq.~\eqref{les} involving integral kernels). 

At this point we take a digression to simplify Eq.~\eqref{ncorr} and write it in  Fourier space. Let $\psi_q^L(\alpha)$ and $\lambda_q^L$ be the single-particle eigenvectors and eigenvalues of the left reservoir Hamiltonian, $\mathcal{H}^L$. Using this and the fact that the left bath is initially described by a grand canonical ensemble with temperature $T_L$ and chemical potential $\mu_L$ we get
\begin{align}
[g_L^+(t-t_0)]_{\nu\sigma}=-i&\theta(t-t_0)\sum_q\psi^{L}_q(\nu) \psi^{L*}_q(\sigma)e^{-i\lambda_q^L(t-t_0)},\\
\expval{c_\nu^\dagger(t_0)c_\sigma(t_0)}&=\sum_q \psi^{L*}_q(\nu)\psi^L_q(\sigma)f_L(\lambda^L_q),\label{bathcorr}
\end{align}
where $f_L(\lambda_q^L)=f(\lambda^L_q,\mu_L,T_L)$ is the Fermi-Dirac distribution function. Using these two equations in Eq.~\eqref{ncorr} we have:
\begin{align}
&\notag\expval{\eta_l^{L\dagger}(t)\eta_m^L(t^\prime)}\\&\hspace{0.2cm}=\sum_{\alpha\nu}V^{L*}_{l\alpha}\left(\sum_q \psi^{L*}_q(\alpha)\psi^L_q(\nu)e^{i\lambda_q^L(t-t^\prime)}f_L(\lambda^L_q)\right) V_{\nu m}^{L^T}\label{rcorr}.
\end{align}
Defining the Fourier transform
\begin{equation}
\tilde{\eta}_l(\omega)=\int_{-\infty}^\infty
\frac{dt}{2\pi} \eta_l(t)e^{i\omega t},
\end{equation}
we finally get the Fourier transform form of Eq.~\eqref{rcorr} as:
\begin{align}
\expval{\tilde{\eta}_l^{L\dagger}(\omega)\tilde{\eta}_m^L(\omega^\prime)}=\Gamma^L_{ml}(\omega)f_L(\omega)\delta(\omega-\omega^\prime)\label{bcorr1},
\end{align}
where $\Gamma_{ml}^L(\omega)=(V^L\rho^LV^{L\dagger})_{ml}$ and $\rho^L_{\alpha\nu}=\sum_q \psi^{L}_q(\alpha)\psi^{L*}_q(\nu)\delta(\omega-\lambda_q^L)$. Using Eq.~\eqref{bcorr1} we can also show that 
\begin{align}
\expval{\tilde{\eta}_l^{L}(\omega)\tilde{\eta}_m^{L\dagger}(\omega^\prime)}=\Gamma^L_{lm}(\omega)\left[1-f_L(\omega)\right]\delta(\omega-\omega^\prime).\label{bcorr2}
\end{align}
The correlation properties of the right bath would be of the same form.

Let us now return back to Eq.~\eqref{les} and obtain its  steady state solution.
 For this we assume that one has taken the limits of infinite bath degrees of freedom and the time $t_0 \to -\infty$. Then it is expected that a steady state should exist provided certain conditions are satisfied~\cite{dhar2006}. For now we assume the existence of a steady state and will re-visit this question in the next section. The Langevin equation in Eq.~\eqref{les} is then amenable to a solution  by Fourier transforms. To this end, we define
\begin{equation}
\tilde{c}_l(\omega)=\int_{-\infty}^\infty
\frac{dt}{2\pi} c_l(t)e^{i\omega t},\label{ftso}\\
\end{equation}
and substitute this in Eq.~\eqref{les} to get
\begin{equation}
[\Pi(\omega)]_{lm}\tilde{c}_m(\omega)-K_{lm}\tilde{c}_m^\dagger(-\omega)=\tilde{\eta}_l^L(\omega)+\tilde{\eta}_l^R(\omega), \label{fts}
\end{equation}
where
\begin{align}
\Pi(\omega)&=\omega-H^W-\Sigma^+_L(\omega)-\Sigma^+_R(\omega),\\
\Sigma_{L}^+(\omega)&=V^{L}g_L^+(\omega)V^{L\dagger},~~\Sigma_{R}^+(\omega)=V^{R}g_R^+(\omega)V^{R\dagger}.
\end{align}
With some algebra one can also show:
\begin{align}
\Gamma_L(\omega)=\frac{1}{2\pi i}\big[\Sigma_{L}^-(\omega)-\Sigma_{L}^+(\omega)\big]\label{gLs},\\\label{gRs} \Gamma_R(\omega)=\frac{1}{2\pi i}\big[\Sigma_{R}^-(\omega)-\Sigma_{R}^+(\omega)\big],
\end{align}
where $\Sigma_{L}^-=[\Sigma_{L}^+]^\dagger$ and $\Sigma_{R}^-=[\Sigma_{R}^+]^\dagger$. We now write  Eq.~\eqref{fts}  in matrix form as:
\begin{equation}
\Pi(\omega)\tilde{C}(\omega)-K\tilde{C}^\dagger(-\omega)=\tilde{\eta}^L(\omega)+\tilde{\eta}^R(\omega), \label{le1}
\end{equation}
where $\tilde{C}(\omega)$, $\tilde{C}^\dagger(\omega)$  and $\tilde{\eta}^{L/R}(\omega)$ are column matrices with components $\tilde{c}_m(\omega)$, $\tilde{c}_m^\dagger(\omega)$ and $\tilde{\eta}_l^{L/R}(\omega)$ respectively. A  complex conjugation of Eq.~\eqref{fts} and transforming $\omega\rightarrow -\omega$  gives us the following  matrix equation:
\begin{equation}
\Pi^*(-\omega)\tilde{C}^\dagger(-\omega)-K^*\tilde{C}(\omega)=\eta^{L\dagger}(-\omega)+\eta^{R\dagger}(-\omega)\label{le2}
\end{equation}
Using Eq.~\eqref{le2} and Eq.~\eqref{le1} we finally obtain the following expression for $\tilde{c}_m(\omega)$:
\begin{align}
\notag\tilde{c}_m(\omega)&=[G_1^+(\omega)]_{ml} \left[\tilde{\eta}_l^L(\omega)+\tilde{\eta}_l^R(\omega)\right]\\&+[G_2^+(\omega)]_{ml} \left[\tilde{\eta}_l^{L\dagger}(-\omega)+\tilde{\eta}_l^{R\dagger}(-\omega)\right],\label{ngef}
\end{align}
where
\begin{align}
G_1^+(\omega)&=\frac{1}{\Pi(\omega)+K[\Pi^*(-\omega)]^{-1}K^\dagger},\label{G1}\\
G_2^+(\omega)&=G_1^+(\omega)K[\Pi^*(-\omega)]^{-1}.\label{G2}
\end{align}
Thus we have obtained the steady state solution in terms of these two nonequilibrium Green's functions. 

\section{Nonequilibrium steady state properties}
\label{sec:ness}
Using the  solution for $\tilde{c}_m (\omega)$ and the noise properties obtained in the previous section, we now proceed to compute expectation values of various physical observables which are  along quadratic functions of the fermionic operators.

\subsection{Steady state particle and energy currents}
 We first  define the particle current in the wire.  Clearly, the rate of change of total number of particles in the left bath, $N_L=\sum_\alpha c_\alpha^\dagger c_\alpha$, gives the particle current, $J_L$, entering the wire from the left reservoir. A straightforward calculation then gives
\begin{align}
&J_L=2\sum_{m\alpha}\Im[V_{m\alpha}^L\expval{c_m^\dagger(t) c_{\alpha}(t)}]\label{currfr}\\&=2\Im\Bigg[\sum_m\int_{-\infty}^\infty\int_{-\infty}^\infty d\omega d\omega^\prime e^{i(\omega-\omega^\prime) t}\notag\\&\hspace{3cm}\expval{c_m^\dagger(\omega)\sum_{\alpha}V^L_{m\alpha}c_\alpha(\omega^\prime)}\Bigg]\label{currfr1}
\end{align}
From the Fourier transform of  Eq.~\eqref{hebl} we have
\begin{equation}
\sum_{\alpha}V_{m\alpha}^L\tilde{c}_\alpha(\omega^\prime)=\eta_m^L(\omega^\prime)+[\Sigma_L^+(\omega^\prime)]_{ml}\tilde{c}_l(\omega^\prime). \label{ftb}
\end{equation}
Using   Eqs.~(\ref{gLs},\ref{gRs},\ref{ngef}) and the correlation properties of the noise terms we finally obtain the following expression for current in the units where $e=h=1$:
\begin{align}
\notag &J_L=\int_{-\infty}^{\infty}d\omega \bigg( T_1(\omega)(f_L^e(\omega)-f_R^e(\omega))\\&+T_2(\omega)(f_L^e(\omega)-f_R^h(\omega))+T_3(\omega)(f_L^e(\omega)-f_L^h(\omega))\bigg),
\label{currexp}
\end{align}
where $G_1^-(\omega)=[G_1^+(\omega)]^\dagger$, $G_2^-(\omega)=[G_2^+(\omega)]^\dagger$ and
\begin{align}
	T_1(\omega)&=4\pi^2\Tr[G_1^+(\omega)\Gamma_R(\omega)G_1^-(\omega)\Gamma_L(\omega)],\\
	T_2(\omega)&=4\pi^2\Tr[G_2^+(\omega)\Gamma_R^T(-\omega)G_2^-(\omega)\Gamma_L(\omega)]~\text{and}\\
	T_3(\omega)&=4\pi^2\Tr[G_2^+(\omega)\Gamma_L^T(-\omega)G_2^-(\omega)\Gamma_L(\omega)].
\end{align}
 We have introduced electron and hole occupation numbers as $f_x^e(\omega)=f(\omega,\mu_x,T_x)$ and $f_x^h(\omega)=f(\omega,-\mu_x,T_x)$, $(x=L,R)$.   
The details of the calculation are presented in the appendix A. A similar expression can be obtained for $J_R$ which we define as the current from the right reservoir into the system.

For $\Delta=0$ case, it is straightforward to see that Eq.~\eqref{currexp} agrees with the expression for the current  obtained in Ref.~[\onlinecite{dhar2006}]. Also, for $\mu_L=-\mu_R=\mu$ and $T_L=T_R=T$ it reduces to 
\begin{align}
\notag &J_L=\notag\int_{-\infty}^{\infty}d\omega [T_1(\omega)+T_3(\omega)](f^e(\omega)-f^h(\omega)),
\end{align}
 This form agrees with the current expression derived in Ref.~[\onlinecite{roy2012}]  for a 1-D Kitaev chain with nearest neighbour interactions. 

 From   Eq.~\eqref{currexp} we see that for $T_L=T_R, \mu_L=\mu_R$,  $J_L, J_R \neq 0$ whenever $\Delta \neq 0$ and, in general, the current at the left end and the right end are different, \emph{i.e}  $J_L \neq -J_R$. This  result initially appears to be surprising, but is basically  due to the fact that the superconducting pairing matrix $\Delta_{lm}$ in the Kitaev wire is not calculated self-consistently but  is taken as a fixed parameter of the wire Hamiltonian.  This becomes  clear if we consider the equation for the total number operator of the wire:
 \begin{align}
\dv{t}(\sum_{l} \langle c_l^\dagger(t) c_l(t)\rangle)=J_S+J_L+J_R,\label{conseq}
 \end{align}
where $J_S=\sum_{l,m}2\Im{K_{lm}\langle c_l^\dagger c_m^\dagger\rangle}$ is the extra contribution from the superconducting terms of the wire Hamiltonian. In the steady state the left hand side vanishes and the fact that $J_L+J_R \neq 0$ can be understood in terms of the extra pairing current $J_S$.  
Physically our set-up corresponds to a wire that is in contact with a superconducting wire and the so-called proximity effect induces superconductivity in the wire. The superconducting substrate  acts as an electron reservoir~\cite{doornenbal2015conductance,lobos2015,akhmerov2011} and acts like a ground for the wire.
Thus current can enter the wire through the left and right reservoirs and flow into the superconductor.
 Also, $J_S$ need not vanish even when the baths are initially at the same chemical potentials and temperatures and hence, $J_L$ and $J_R$ may take  non-zero values.  Note that imposing the self-consistency condition, namely 
\begin{align}
K_{lm}= \expval{c_m c_l}= \langle c_l^\dagger c_m^\dagger\rangle^*,
\end{align}
for all $l,m$, would give $J_S=0$ and in that case we would get the expected charge conservation condition $J_L=-J_R$.

	We comment on the physical interpretation of the three different parts in Eq.~\eqref{currexp}:  the first term  corresponds to normal electrons being transmitted from the left to the right bath (normal transmission), the second term  corresponds to the process of an electron from the left bath being scattered as a hole into the  right bath (Andreev transmission) while the third term  corresponds to the electron from the left bath scattered back as a hole into the left bath again (Andreev reflection). The probability of these three processes are then given respectively by $T_1(\omega)$, $T_2(\omega)$ and $T_3(\omega)$.   Defining the conductance at the left end   (in units of $e^2/h$) by 
	\begin{equation}
	G_L(T_L,\mu_L)=\pdv{J_L}{\mu_L},
	\end{equation}
we get at zero temperature ($T_L=T_R=0$):
	\begin{equation}
	G_L=T_1(\mu_L)+T_2(\mu_L)+T_3(\mu_L)+T_3(-\mu_L).\label{GLnegf1}
	\end{equation}
Due to the particle-hole symmetry of the Hamiltonian, we expect $T_2(\omega)$ and $T_3(\omega)$ to be even functions of $\omega$. Therefore, $T_3(\omega)$ contributes twice to the conductance which represents the fact that in Andreev reflection, a total of two electrons are transferred across the junction as a single cooper pair. In Sec.~(\ref{sec:numerics}), we numerically evaluate and plot the three components $T_1(\omega), T_2(\omega), T_3(\omega)$ for the Kitaev chain model and demonstrate the above points.

  We now turn to the computation of the energy current, which is readily obtained using our formalism. The energy current coming into the wire  from the left end can be obtained by the rate of change of left bath Hamiltonian, $\mathcal{H}_L$. So,  we consider $\frac{d}{dt}\expval{\mathcal{H}^L}$ and then use the Heisenberg equations of motion for the left reservoirs operators to obtain:
  \begin{equation}
  	J_L^H=-\frac{d}{dt}\expval{\mathcal{H}^L}=2\sum_{l\nu}\Im{[V^LH^L]_{l\nu}\langle c_l^\dagger c_\nu \rangle},
  \end{equation} 
where $J_L^H$ is the energy current flowing into the wire. This can be simplified  by using Eq.~\ref{hebl} to obtain,
\begin{align}
	J_L^H&=-2\Im\Bigg[\sum_m\int_{-\infty}^\infty\int_{-\infty}^\infty d\omega d\omega^\prime e^{i(\omega-\omega^\prime) t}\notag\\&\hspace{3cm}\omega^\prime \expval{c_m^\dagger(\omega)\sum_{\alpha}V^L_{m\alpha}c_\alpha(\omega^\prime)}\Bigg].
\end{align}
Comparing this with the expression for particle current in Eq.~\eqref{currfr1}, it can be seen that this would yield the same expression with an extra factor of $\omega$ in the integral. After some  simplification this then gives:
\begin{align}
 &J_L^H=\int_{-\infty}^{\infty}d\omega ~\omega \left[T_1(\omega)+T_2(\omega)\right](f_L^e(\omega)-f_R^e(\omega)).
\label{ecurrexp}
\end{align}

The low temperature thermal conductance is given by
\begin{equation}
	G^H_L=\frac{dJ_L^H}{dT_L}=\frac{k_B^2\pi^2T_L}{6}G_T(\mu_L), \label{econd}
\end{equation}
where, $G_T(\mu_L)=2(T_1(\mu_L)+T_2(\mu_L))$ and is of similar form as obtained in Ref.~\onlinecite{bondyopadhaya2020nonequilibrium}. Therefore, only two processes contribute to the heat current the normal transmission and the Andreev transmission. The Andreev reflection term does not contribute to the energy current since the particle and hole each carry a unit of energy and so no net energy is transferred across the junction. Note also that, unlike the electron current,  the energy current is the same at both ends of the wire and indeed anywhere inside the wire since energy is conserved. 

   The general expressions for particle current and energy current  in Eqs.~(\ref{currexp},\ref{ecurrexp}) are two of our main results. These expressions provide  compact formulas for  thermal and particle conductances and may be used to study  these physical quantities in systems defined on arbitrary  lattices with interactions between arbitrary sites.

 \subsection{Two point correlations}
 We now compute the full two-point correlation matrices $\la c^\dagger_l c_m \ra,~\la c_l c_m \ra,~\la c^\dagger_l c^\dagger_m \ra$ in the NESS. These would allow one to obtain the local particle densities, $\la c^\dagger_l c_l \ra$, and local currents (normal and superconducting) anywhere in the system.  
We start by writing the steady state correlations in the Fourier representation:
 \begin{align}
 N_{lm}^{SS}=\expval{c_m^\dagger(t)c_n(t)}=\int \int d\omega d\omega^\prime e^{i(\omega^\prime-\omega)t}\expval{c_m^\dagger(\omega)c_n(\omega^\prime)}.
 \end{align}
Then using the solution in Eq.~\eqref{ngef} and the noise properties a straightforward computation gives:
 \begin{align}
 &N_{lm}^{SS}=\sum_{x=L,R}\int d\omega\bigg[[G_1^+(\omega)\Gamma_x(\omega)G_1^-(\omega)]_{nm}f_x^e(\omega)\notag\\&\hspace{1.5cm}+[G_2^+(\omega)\Gamma^T_x(-\omega)G_2^-(\omega)]_{nm} f_x^h(\omega)\bigg].\label{denexp}
 \end{align}
A  similar computation gives
 \begin{align}
&M_{lm}^{SS}=\expval{c_i(t)c_j(t)}\notag\\&=\sum_{x=L,R}\int d\omega [Q_x(\omega)]_{ij}+[Q_x^T(\omega)-Q_x(\omega)]_{ij}f_x(\omega),\label{cccorr}
 \end{align}
where $Q_{L/R}(\omega)=G_1^+(\omega)\Gamma_{L/R}(\omega)G_2^{+T}(-\omega)$. Substituting this in the expression for $J_S=\sum_{l,m}2\Im{K_{lm}\langle c_l^\dagger c_m^\dagger\rangle}$, we get its steady state value,
 \begin{align}
 	\notag J_S=2\int d\omega &\Im{\Tr[Q_L^\dagger(\omega) K]}(2f_L(\omega)-1)\\&+\Im{\Tr[Q_R^\dagger(\omega) K]}(2f_R(\omega)-1)
 \end{align}
 
From Eq.~\eqref{cccorr}, it also follows that  $\expval{\acomm{c_i(t)}{c_j(t)}}=\int d\omega [Q_L(\omega)  + Q_L^T(\omega) +Q_R(\omega)+Q_R^T(\omega)]_{ij}=I_{ij} $. It turns out that these integrals do not always vanish. This is at first surprising since we expect that the usual anti-commutation properties of the fermionic operators should hold. The underlying reason is  that  the results presented so far assume the existence of a steady state.   However this is true only if there are no bound states in the system (wire+baths). In case there are bound states present in the system, then their contributions  to the expressions of the correlations  have to be added separately. For the case of Eq.~\eqref{cccorr},  the contribution from the bound state  would  ensure the vanishing of  $\expval{\acomm{c_i(t)}{c_j(t)}}$.  To calculate the contributions of the bound states  requires one to use an approach involving the diagonalization of the  full Hamiltonian of the two baths and wire and identification of the bound states. We describe this in the next section where we present the contributions of the bound states to the correlations. We  relegate the details of the calculation to  an appendix.~\ref{appendix2}. In section.~\ref{sec:numerics} we will demonstrate numerically that these contributions ensure that the commutation relations hold and also show their effect on the density correlations. 

\section{An exact numerical approach for computing correlations in finite systems and the bound state contribution to the two point correlators} 
\label{sec:timeevol}
The fact that our system is described by a quadratic Hamiltonian means that the exact diagonalization of the system becomes a much simpler problem~\cite{blaizot1986,roy2019}. Let $N_S$ be the total number of lattice sites in the entire system of wire and the two reservoirs. Then instead of diagonalizing a   $2^{N_S} \cross 2^{N_S}$ matrix, the problem reduces to the diagonalization of a $2 N_S \cross 2 N_S$ matrix.  To see this we define a $2 N_S$-component column vector: 
\begin{equation}
	\chi=\begin{pmatrix} C \\ C^\dagger \end{pmatrix}, \rm{where}~~ C=\begin{pmatrix} C_W \\C_L \\ C_R \end{pmatrix},
\end{equation}
and $C_W$, $C_L$ and $C_R$ are column vectors containing the wire, left bath and the right bath operators  respectively. Note that $\chi_{N_S+i}^\dagger=\chi_i$, for $i=1,2,\ldots,N_S$. We can then write the Hamiltonian in Eq.~\eqref{hamil} in the  form 
\begin{align}
	\mathcal{H}=\frac{1}{2}\chi^\dagger \mathcal{Z} \chi + \frac{1}{2} \Tr[H_S]\label{Zhamil},
\end{align}
where $\mathcal{Z}$ is a $2 N_S \cross 2 N_S$ matrix defined as
\begin{equation}
	\mathcal{Z}=\begin{pmatrix}
	H_S && {K_S} \\
	{K_S^\dagger} && -H_S^*\end{pmatrix},
\end{equation} 
with 
 \begin{align}
H_S=\begin{pmatrix}H_W && V_L && V_R\\ V_L^\dagger && H_L&& 0\\ V_R^\dagger && 0 && H_R\end{pmatrix}\hspace{0.1cm}\text{and}\hspace{0.1cm}
{K_S}=\begin{pmatrix}K&& 0&& 0\\ 0&&0&& 0\\ 0 && 0 && 0\end{pmatrix}.
\end{align}
As can be easily verified, the $2N_S$ eigenvectors of the matrix $\mathcal{Z}$ occur in pairs of the form
 \begin{equation}
 	\psi_{i}=\begin{pmatrix}
 	u_1(\epsilon_i)\\u_2(\epsilon_i) \\ . \\. \\u_{N_S}(\epsilon_i) \\v_1(\epsilon_i) \\. \\. \\v_{N_S}(\epsilon_i)
 	\end{pmatrix} \hspace{0.5cm} \phi_{i}=\begin{pmatrix}
 	v_1^*(\epsilon_i)\\v_2^*(\epsilon_i) \\ . \\. \\v_{N_S}^*(\epsilon_i) \\u_1^*(\epsilon_i) \\. \\. \\u^*_{N_S}(\epsilon_i) 	
 	\end{pmatrix},~~i=1,2,\ldots,N_S ,\label{evecs}
 \end{equation}
where the eigenvectors $\psi_i$ and $\phi_i$ correspond respectively to eigenvalues  $\epsilon_i$ and $-\epsilon_i$. Let us define the $N_S \cross N_S$ matrices $U$, $V$ and $E$ with matrix elements $U_{si}=u_s(\epsilon_i)$, $V_{si}=v_s(\epsilon_i)$ and $E_{ij}=\epsilon_i \delta_{ij}$, respectively.
Then we see that the matrix $W$ which diagonalizes $\mathcal{Z}$ has the structure
\begin{equation}
	W=\begin{pmatrix}
	U && V^* \\
	V && U^*\end{pmatrix}, \label{Wstruct}
\end{equation} 
so that
\begin{equation}
	W^\dagger Z W =\begin{pmatrix}
	E && 0 \\
	0 && -E \end{pmatrix}.
\end{equation} 
 We define new fermionic variables $\zeta=W^\dagger\chi$ and note that due to the structure in Eq.~\eqref{Wstruct}, $\zeta_{N_S+i}=\zeta_i^\dagger$, for $i=1,2,\ldots,N_S$. Note that this transformation mixes the operators corresponding to different sites of the wire and the bath, and the index $i$ does not refer to any lattice site. The $\zeta_i$ correspond to the ``normal modes'' of the system. In this basis the  Hamiltonian then takes the form
  \begin{equation}
  	\mathcal{H}=\sum_{i=1}^{N_S} \epsilon_i\left(\zeta_i^\dagger(t)\zeta_i(t) -\frac{1}{2}\right)+\frac{1}{2} \Tr [H_S].
  \end{equation}
 The evolution of the $\zeta$ operators is simply given by $\zeta_j(t)=e^{-i\epsilon_j t}\zeta_j(0)$. Therefore, a two point correlator of the original operators at any time $t$ can be expressed in terms of $\zeta$ operators at $t=0$ via the transformation $W$. For the correlator $\langle c_p^\dagger(t) c_q(t) \rangle$, where $p,q$ denotes any site on the entire system,  we thus obtain:
 \begin{align}
&\langle c_p^\dagger(t) c_q(t) \rangle=\sum_{l,m=1}^{N_S}\bigg[\mathcal{T}_{{N_S}+p,q}^{{N_S}+l,{N_S}+m}e^{i(\epsilon_l+\epsilon_m)t}\langle\zeta_l^\dagger\zeta_m^\dagger\rangle\notag\\&+\mathcal{T}_{{N_S}+p,q}^{lm}e^{-i(\epsilon_l+\epsilon_m)t}\expval{\zeta_l\zeta_m}+\mathcal{T}_{{N_S}+p,q}^{{N_S}+l,m}e^{-i(-\epsilon_l+\epsilon_m)t}\langle\zeta_l^\dagger\zeta_m\rangle\notag\\&+\mathcal{T}_{{N_S}+p,q}^{l,{N_S}+m}e^{-i(\epsilon_l-\epsilon_m)t}\langle\zeta_l\zeta_m^\dagger\rangle\bigg],
\label{corrnum}
 \end{align} 
where $\mathcal{T}_{pq}^{lm}=W_{pl}W_{qm}$ and $\zeta_i$ in the above equation denotes $\zeta_i(0)$. Using the  transformation $\zeta=W^\dagger \chi$, the two point correlations of the $\zeta$ operators at $t=0$ can be determined from the two point correlations of $c_p$ and $c_p^\dagger$ at $t=0$, which are  known once  the initial state of the system is specified. In particular we know these correlations for the product initial state used in the previous section, where the reservoirs are described by thermal states  with specified temperatures and chemical potentials, while the system is in an arbitrary initial state.  

The numerical approach thus consists of finding the eigenspectrum of the matrix $\mathcal{Z}$ and then computing the time evolution of any two-point correlator using Eq.~\eqref{corrnum}.  Our interest will be in looking at correlations in the wire. For a finite bath we expect to see steady state behaviour of the wire correlations in a time window, which is after some initial transients and before 
the finite bath effects show up. Thus the correlations would first show some initial evolution, then show a long plateau before finite size effects show up. The steady state properties can be extracted from the plateau region 
 We will  use this procedure in the next section  to directly verify the  steady state results given by the analytic expressions in the previous section. 
 
{\bf Bound states:} As discussed earlier we can look for the existence of bound states by examining the spectrum of the matrix $\mathcal{Z}$.  The bound state corresponds to  states  which lie outside the band width of the baths and the corresponding eigenvector would be spatially localized. The existence of such bound states in general cause persistent oscillations and  steady state properties can become periodic in time. We are in fact  now in a position to write down the bound state contributions to Eq.~\ref{denexp} and Eq.~\ref{cccorr} respectively. Let us write the  eigenvector of the matrix $\mathcal{Z}$ corresponding to a bound state with eigenvalue $E$ in the form  $\begin{pmatrix} \Psi_{E} \\ \Phi_{E} \end{pmatrix}$   where $\Psi_{E}(q)=u_q(E)$ and $\Phi_{E}(q)=v_q(E)$($q$ runs from $1$ to $N_S$). Following Ref.~[\onlinecite{dhar2006}], we then see that  the contribution of the bound states to  Eq.~\ref{denexp} and Eq.~\ref{cccorr} are given by
\begin{widetext}
\begin{align}
N_{lm}^{BS}(t)=&\sum_{i,j,E_b,E_{b^\prime},x=L,R}e^{i(E_{b^\prime}-E_b)t}\Bigg[\Psi_{E_b}(l)\Psi_{E_b}^*(j)\Psi_{E_{b^\prime}}(i)\Psi_{E_{b^\prime}}^*(m)\int d\omega \frac{[\Gamma_x(\omega)]_{ji}f_x(\omega)}{(\omega-E_b)(\omega-E_{b^\prime})}\notag\\&+\Psi_{E_b}(l)\Phi_{E_b}^*(j)\Phi_{E_{b^\prime}}(i)\Psi_{E_{b^\prime}}^*(m)\int d\omega\frac{[\Gamma_x^T(\omega)]_{ji}(1-f_x(\omega))}{(\omega+E_b)(\omega+E_{b^\prime})}+\Psi_{E_b}(l)\Phi_{E_b}^*(j)\Phi_{E_{b^\prime}}(i)\Psi_{E_{b^\prime}}^*(m)\delta_{ij}\Bigg]\label{Nlm},~~\text{and}\\
M_{lm}^{BS}(t)=&\sum_{i,j,E_b,E_{b^\prime},x=L,R}e^{-i(E_{b^\prime}+E_b)t}\Bigg[\Psi_{E_b}(l)\Psi_{E_b}^*(j)\Phi_{E_{b^\prime}}(i)\Psi_{E_{b^\prime}}^*(m)\int d\omega \frac{[\Gamma_x(\omega)]_{ji}f_x(\omega)}{(\omega-E_b)(\omega+E_{b^\prime})}\notag\\&+\Psi_{E_b}(l)\Phi_{E_b}^*(j)\Psi_{E_{b^\prime}}(i)\Psi_{E_{b^\prime}}^*(m)\int d\omega\frac{[\Gamma_x^T(\omega)]_{ji}(1-f_x(\omega))}{(\omega+E_b)(\omega-E_{b^\prime})}+\Psi_{E_b}(l)\Psi_{E_b}^*(j)\Phi_{E_{b^\prime}}(i)\Psi_{E_{b^\prime}}^*(m)\delta_{ij}\Bigg]\label{Mlm}
\end{align}
\end{widetext}
respectively. The sum in these expressions runs over all bound state eigenvectors of $\mathcal{Z}$ with  positive as well as negative eigenvalues. These are  identified from  the spectrum of $\mathcal{Z}$ as eigenvectors corresponding to eigenvalues  which  lie outside the band. In the next section, we demonstrate numerically the fact that the addition of these two corrections to the steady state values gives us the exact long time behaviour of the correlators of the wire. As is clear from these expressions, the bound state contribution would in general cause persistent oscillations in two point correlations of the wire and hence, in all the steady state properties of the wire. The frequencies of these oscillations would be the sum and differences of the energies of the corresponding high energy bound states.

\section{Numerical verification of QLE-NEGF results and the bound state contributions in a Nearest neighbour Kitaev chain}
\label{sec:numerics}
We apply the numerical approach of the previous section on the one-dimensional Kitaev chain to verify the analytical results in Sec.~\eqref{sec:ness}. We consider a one-dimensional system with $N$ sites on the wire and $N_b$ on each of the two baths and so $N_S=N+2 N_b$.  The full system Hamiltonian is given by:
\begin{align}
&\mathcal{H}=\cH_W+\cH_L+\cH_{WL}+\cH_R+\cH_{WR}\notag\\
&=\sum_{j=1}^{N-1} \left[-\mu_wa_j^\dagger a_j+\big(-\eta_w a_j^\dagger a_{j+1}+\Delta a_ja_{j+1}+\text{c.c.}\big)\right]\notag
\\&+\sum_{\alpha=1}^{N_b-1}\left[-\eta_b(b_\alpha^{L\dagger} b_{\alpha+1}^L+b_{\alpha+1}^{L\dagger} b_\alpha^L)\right]-V_L(a_{1}^\dagger b_1^L+b_1^{L\dagger} a_{1})\notag
\\&+\sum_{\alpha'=1}^{N_b-1}\left[ -\eta_b(b_{\alpha'}^{R\dagger} b_{\alpha'+1}^R+b_{\alpha'+1}^{R\dagger} b_{\alpha'}^R)\right]-V_R(a_{N}^\dagger b_1^R+b_1^{R\dagger} a_{N}),\label{nnh}
\end{align}
where $\{a_j,a_j^\dagger\}$, $\lbrace b_\alpha^R,b_\alpha^{R\dagger}\rbrace$,$\lbrace b_{\alpha'}^L,b_{\alpha'}^{L\dagger}\rbrace$ are annihilation and creation  operators on the wire, right and left bath sites respectively. As in Sec.~\eqref{sec:QLE}, we start from the initial state:
\begin{equation}
\rho=\frac{e^{-\beta_L (\cH_L-\mu_L \mathcal{N}_L)}}{Z_L}\otimes\ket{0}\bra{0}\otimes\frac{e^{-\beta_R (\cH_R-\mu_R \mathcal{N}_R)}}{Z_R},
\end{equation}
where $\mathcal{N}_{L,R}$ are the number operators in the baths, $Z_x=\Tr(e^{-\beta_x (\cH_x-\mu_x \mathcal{N}_x)})$, $x=L,R$, the partition functions of the baths and $|0\ra \la 0|$ refers to the wire being initially completely empty.  With this choice of the initial state, we can compute all the $t=0$ correlations required in Eq.~\eqref{corrnum}. The eigenvalues and eigenfunctions of the matrix $\mathcal{Z}$ defined in the previous section, corresponding to the Hamiltonian Eq.\eqref{nnh}, can be easily computed numerically for chains of finite length $N_S$. 
 For our numerical example we take $N=2$, $N_b=100$, and  use Eq.~\eqref{corrnum}  in the previous section to calculate the time evolution,  at any finite time, of the currents $J_L(t)=-2V_L \rm{Im}[ \la a_1^\dagger(t) b_1^L(t) \ra] , J_R(t)=-2V_R \rm{Im}[\la a_N^\dagger(t) b_1^R(t) \ra] $ at the two boundaries and  the densities $N_1(t)= \la a_1^\dagger(t) a_1(t) \ra, N_2(t)= \la a_2^\dagger(t) a_2(t)\ra$. 
\begin{figure}[htb!]
	\centering
	\subfigure[]{
		\includegraphics[width=40mm]{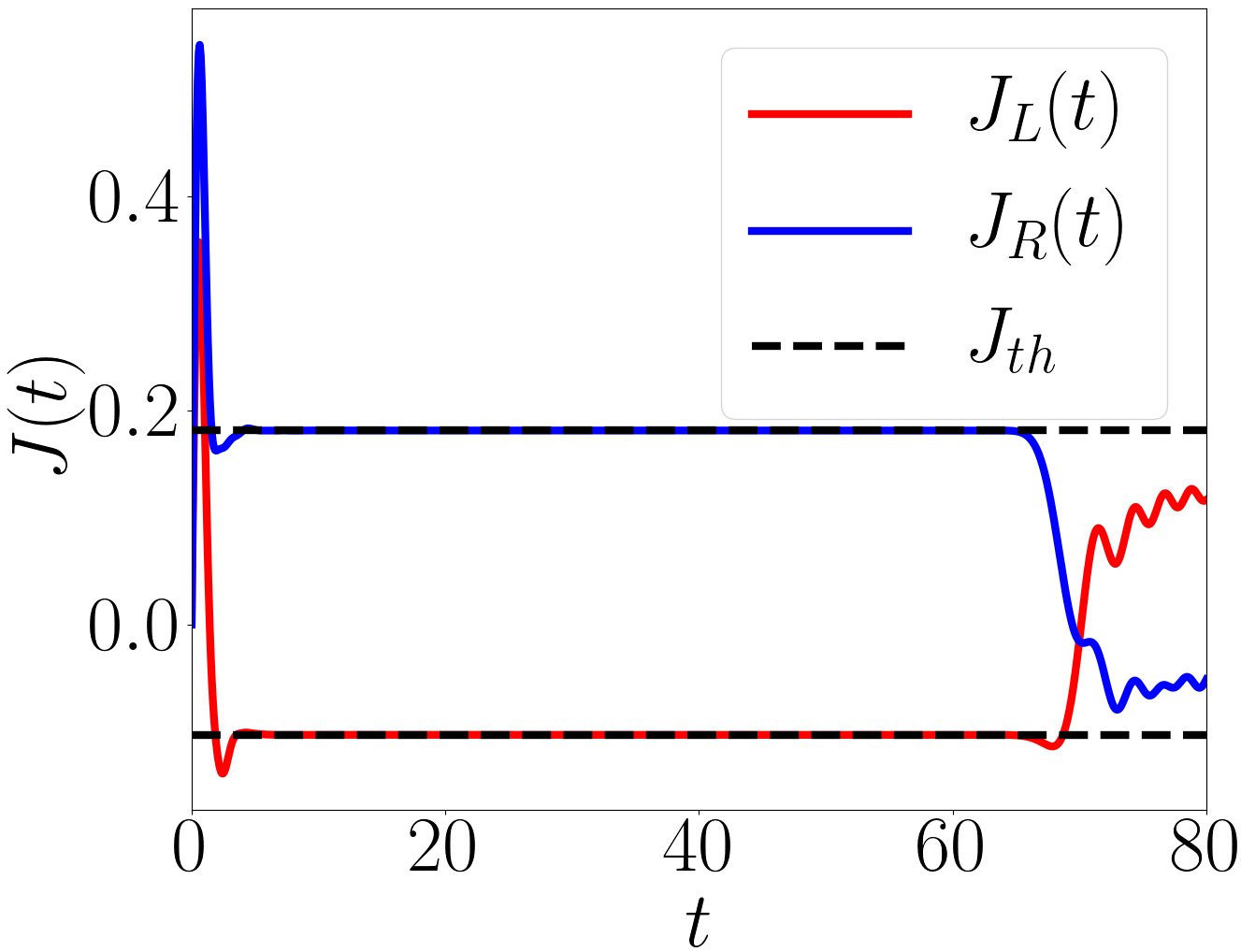}
	}
		\subfigure[]{
		\includegraphics[width=40mm]{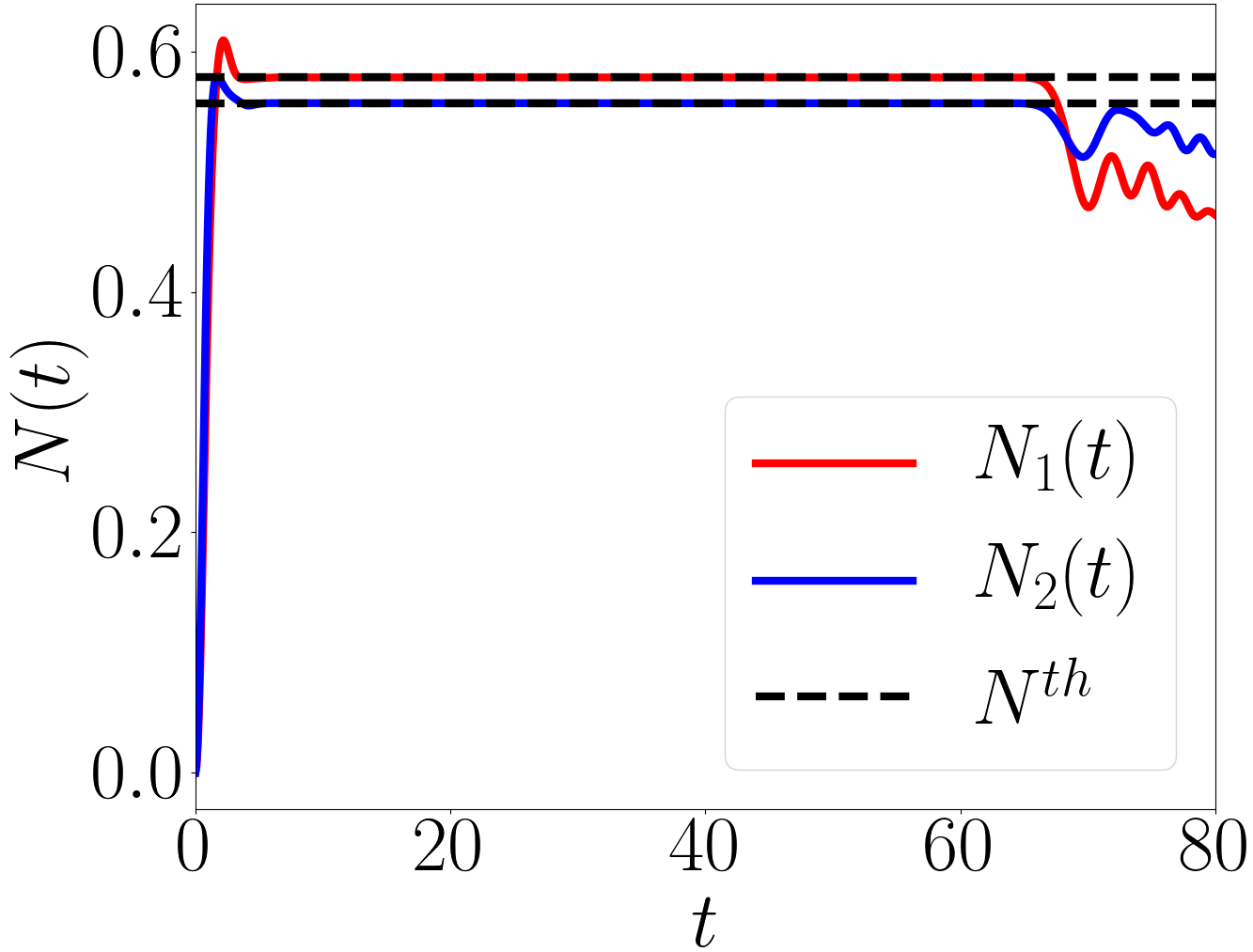}
	}
		\subfigure[]{
		\includegraphics[width=40mm]{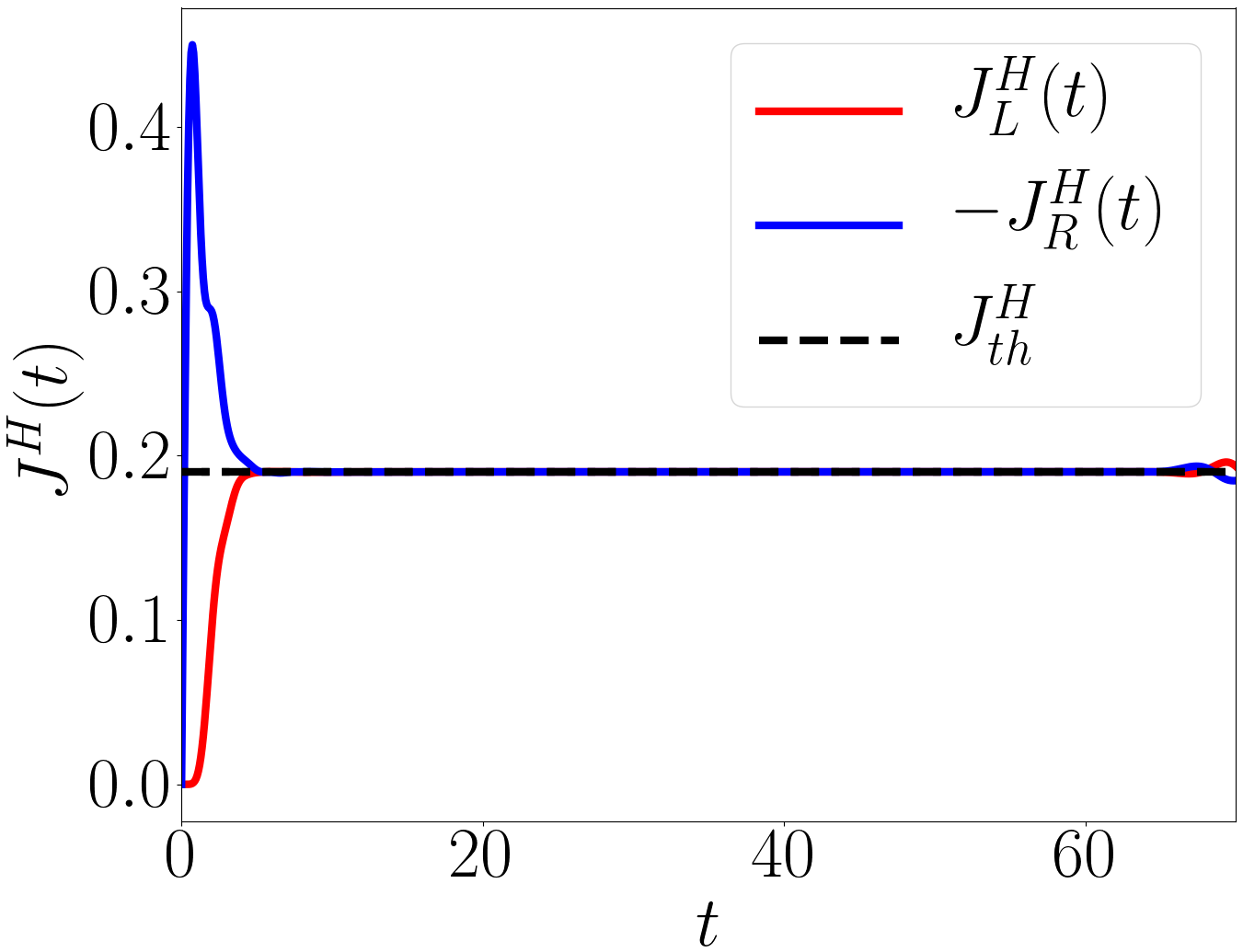}
	}
		\subfigure[]{
		\includegraphics[width=40mm]{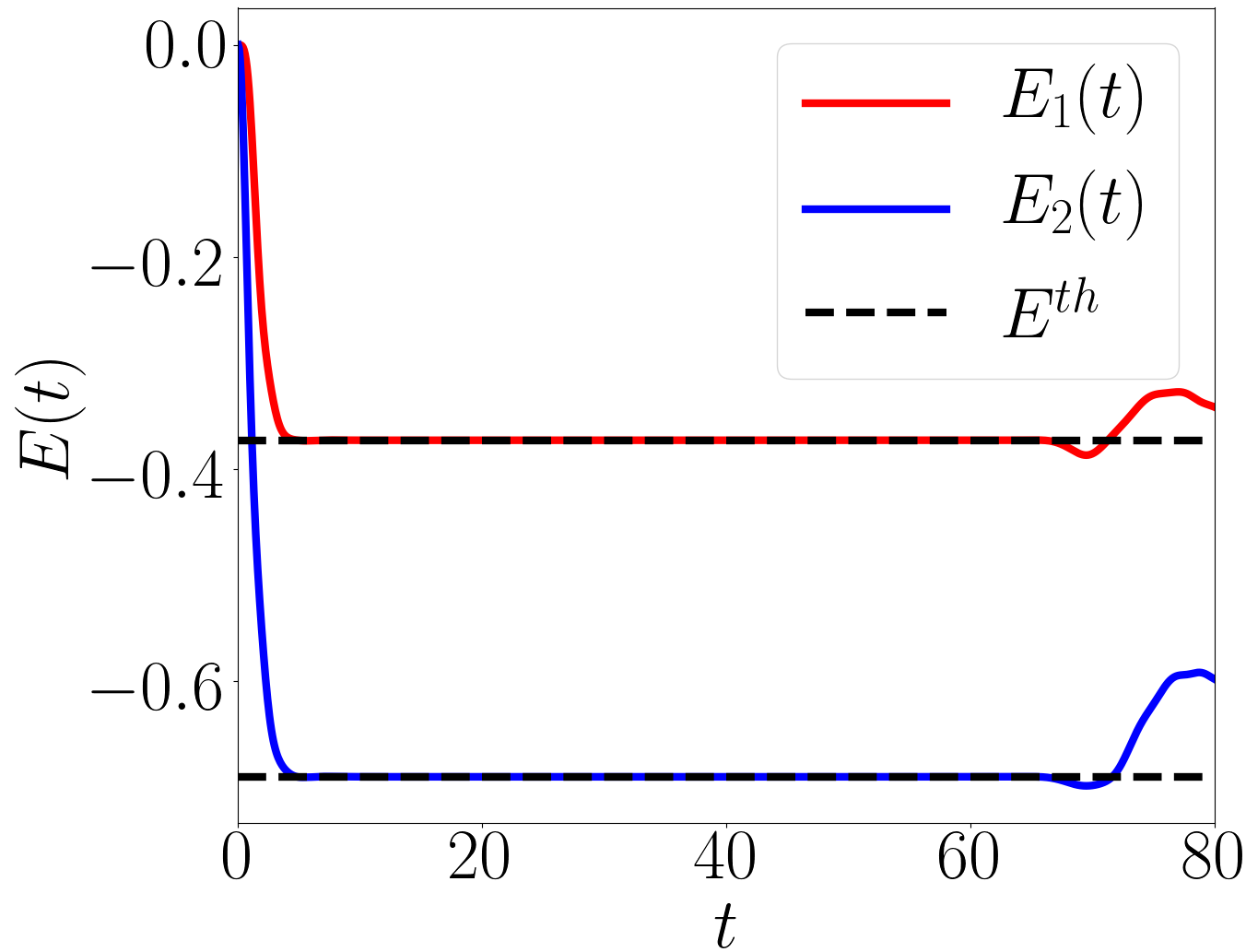}
	}
	\caption{Comparison of numerical time-evolution and analytical steady state results: Parameter values ---  $ N=2$,
		$N_b=100$, $\mu_R=1$, $\beta_R=10$, $\mu_L=0$, $\beta_L=0$, $V_L=V_R=\eta_w=1$, $\mu_w=0$, $\eta_b=1.5$ and  $\Delta=0.40$. 
		  (a) Comparison of the numerically calculated particle current at the left, $J_L(t)$, and the right, $J_R(t)$, end of the wire with the corresponding value, $J_{th}$, given by the expression  in Eq.~\eqref{currexp}. (b) Comparison of the numerically calculated densities, $N_1(t)=\langle a_1^\dagger(t)a_1(t)\rangle$ and $N_2(t)=\langle a_2^\dagger(t)a_2(t)\rangle$, on the two sites of the wire with the corresponding value, $N_{th}$, given by the expression  in Eq.~\eqref{denexp}.  Similarly, (c) and (d) show the comparison of the energy current and the energy density from direct numerics with the  values obtained from steady state expressions. Note that  the left and the right heat currents have the same magnitude unlike the particle currents. The initial oscillations seen in the plots correspond to the transient phase, while  the behaviour near $t=60$ is due to the finite size of the baths. In the intermediate region we see perfect agreement between the numerical solution and the steady state value.}
	\label{currplts}
\end{figure}
 \begin{figure}[htb!]
	\centering
	\subfigure[$\Delta=2$]{
		\includegraphics[width=4.cm,height=35mm]{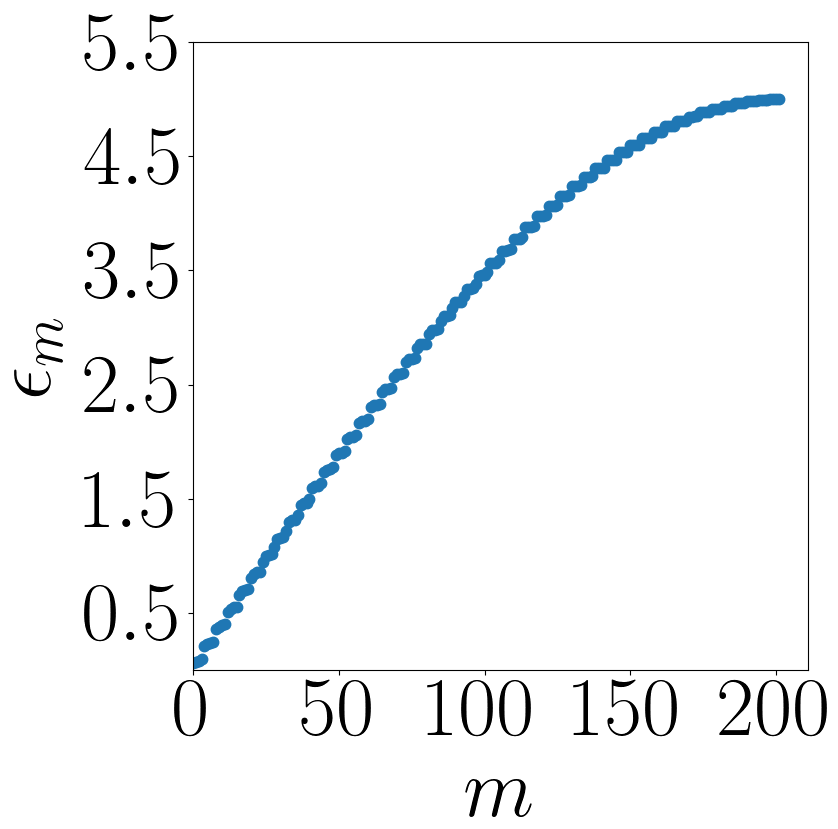}
	}
	\subfigure[$\Delta=5$]{
		\includegraphics[width=4.cm,height=35mm]{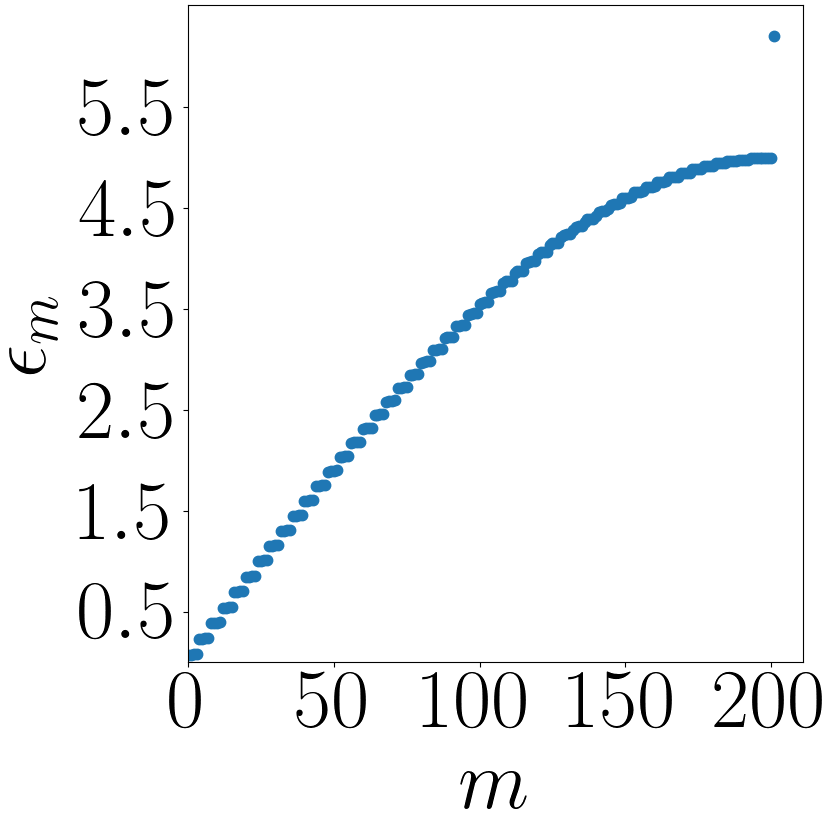}
	}
	\caption{Spectrum of the entire system at parameter values  $ N=2$,
		$N_b=100$,  $V^L=V^R=\eta_w=1$, $\mu_w=0$ and $\eta_b=2.5$ for two values of $\Delta$. In (a) we do not see any discrete energy level while in (b) a discrete energy level outside the main band can be seen.  As discussed in the text, the non-existence of a steady state, indicated in a non-vanishing $I_{ij}$, is related to the existence of the discrete level which corresponds to a bound state (see Fig.~\eqref{Ivar}).}
\label{bounds}
\end{figure}

 The general analytic expressions for the steady state properties of the wire are in terms of the two Green's functions $G_1^+(\omega)$ and $G_2^+(\omega)$ defined in Eq.~\eqref{G1} and Eq.~\eqref{G2} respectively. For the Hamiltonian  given in Eq.~\eqref{nnh}, the various matrices involved in them  take simpler forms  and one finds~\cite{dhar2006}:
\begin{align}
	K_{ij}&= \Delta(\delta_{i,j+1}-\delta_{i,j-1})\label{Kmat},\\
	[\Sigma^+_{L}(\omega)]_{ij}&=V_L^2 g(\omega) \delta_{i1}\delta_{j1},\\
	\pi[\Gamma_{L}(\omega)]_{ij}&=V_L^2 \Im{g(\omega)} \delta_{i1}\delta_{j1},  \\ 
	[\Sigma^+_{R}(\omega)]_{ij}&=V_R^2 g(\omega)\delta_{iN}\delta_{jN},\\
	\pi[\Gamma_{R}(\omega)]_{ij}&=V_R^2 \Im{g(\omega)}\delta_{iN}\delta_{jN},\\
	[\Pi(\omega)]_{ij}&=\omega\delta_{ij}+\eta_s(\delta_{i,j+1}+\delta_{i,j-1}),\notag\\& -V_L^2 g(\omega)\delta_{i1}\delta_{j1}-V_R^2 g(\omega)\delta_{iN}\delta_{jN},
\end{align}
where $g(\omega)=[g^+_L(\omega)]_{11}$. Since $g^+_L(\omega)$ is the inverse of a tri-diagonal matrix, it can be shown that~\cite{dhar2006}
\begin{align}
	 g(\omega)= 
	\begin{cases}
	\frac{1}{\eta_b}\left(\frac{\omega}{2\eta_b}-\sqrt{\frac{\omega^2}{4\eta_b^2}-1}\right),& \text{if } \omega > 2\eta_b\\
	\frac{1}{\eta_b}\left(\frac{\omega}{2\eta_b}+\sqrt{\frac{\omega^2}{4\eta_b^2}-1}\right), & \text{if } \omega < -2\eta_b\\
	\frac{1}{\eta_b}\left(\frac{\omega}{2\eta_b}-i\sqrt{1-\frac{\omega^2}{4\eta_b^2}}\right), & \text{if } \abs{\omega} < 2\eta_b\label{gomega}.
	\end{cases}
\end{align}
 Using Eqs.~(\ref{Kmat}-~\ref{gomega}) we can compute the steady state value of the current and densities by direct substitution of these expressions in Eq.~\eqref{currexp} and Eq.~\eqref{denexp}. The integrations over $\omega$ in the resulting expressions are carried out numerically. 
 In  Fig.~\eqref{currplts} we show the comparison between the  values for the steady state currents (particle and heat) and densities  obtained from Eqs.~(\ref{currexp},\ref{ecurrexp},\ref{denexp}) with the corresponding values obtained from the direct time evolution. 
\begin{figure}[htb!]
	\centering
	\subfigure[]{
		\includegraphics[width=4cm,height=40mm]{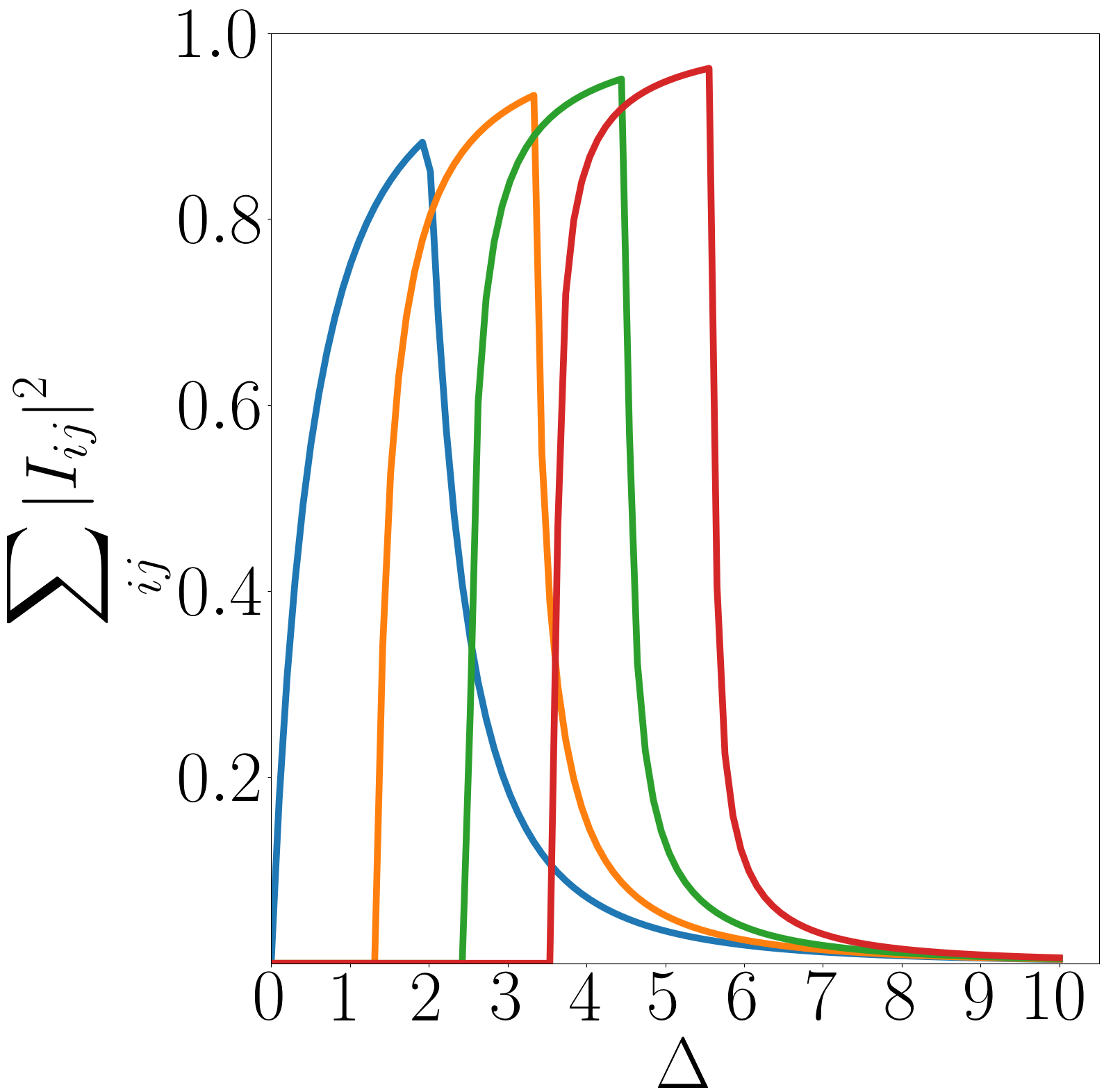}
	}
	\subfigure[]{
		\includegraphics[width=4cm,height=40mm]{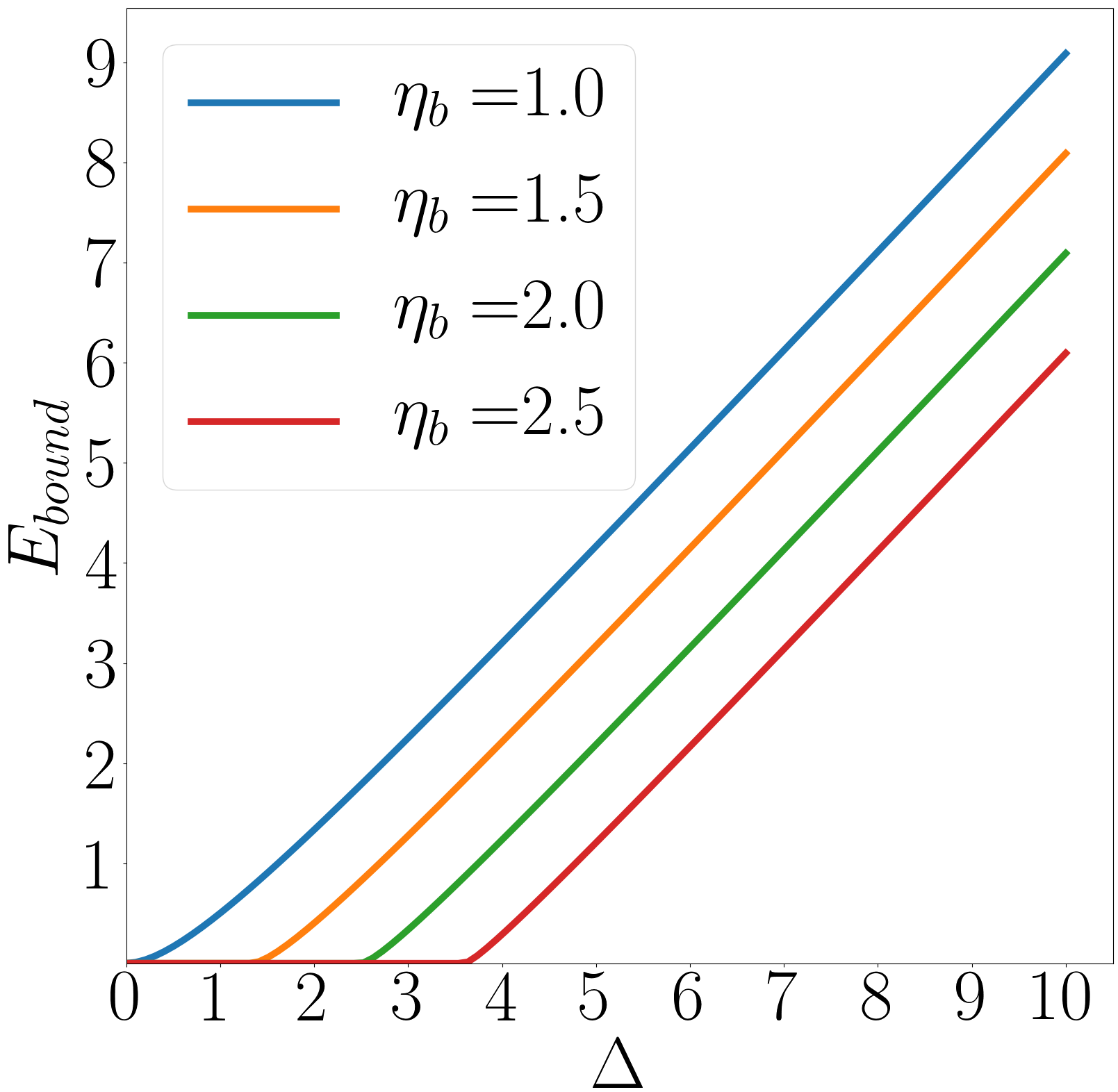}
	}
		\subfigure[]{
		\includegraphics[width=8cm,height=35mm]{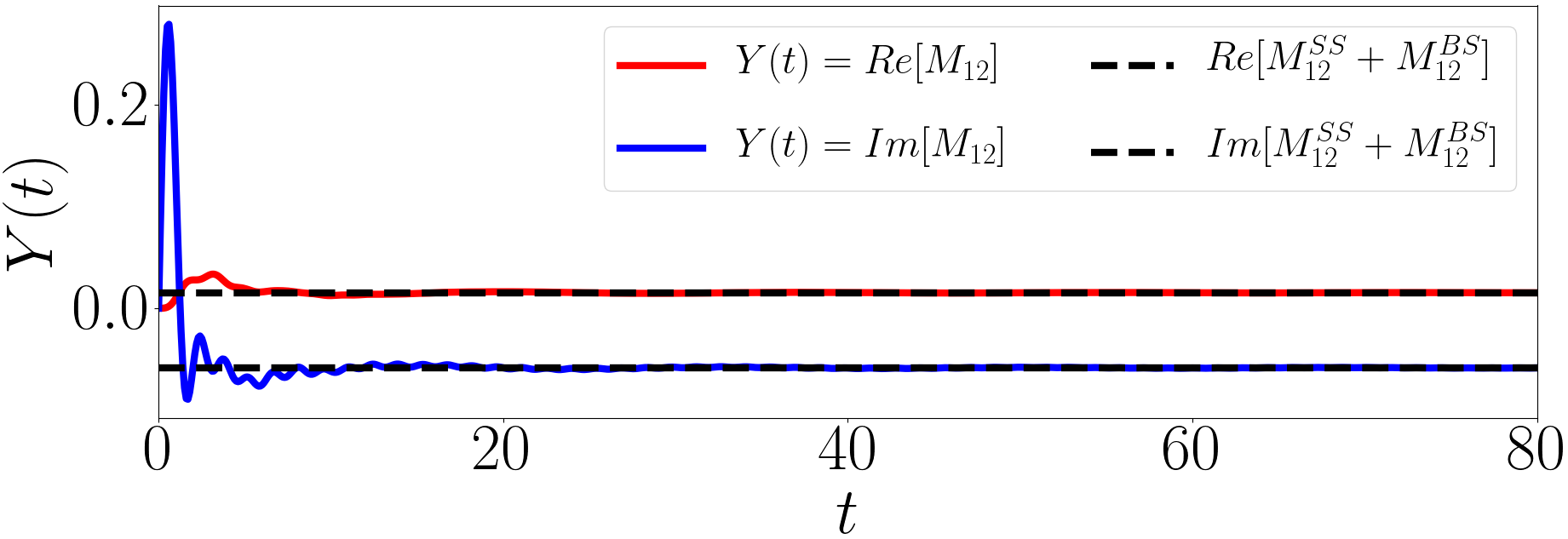}
	}
	\subfigure[]{
		\includegraphics[width=8cm,height=35mm]{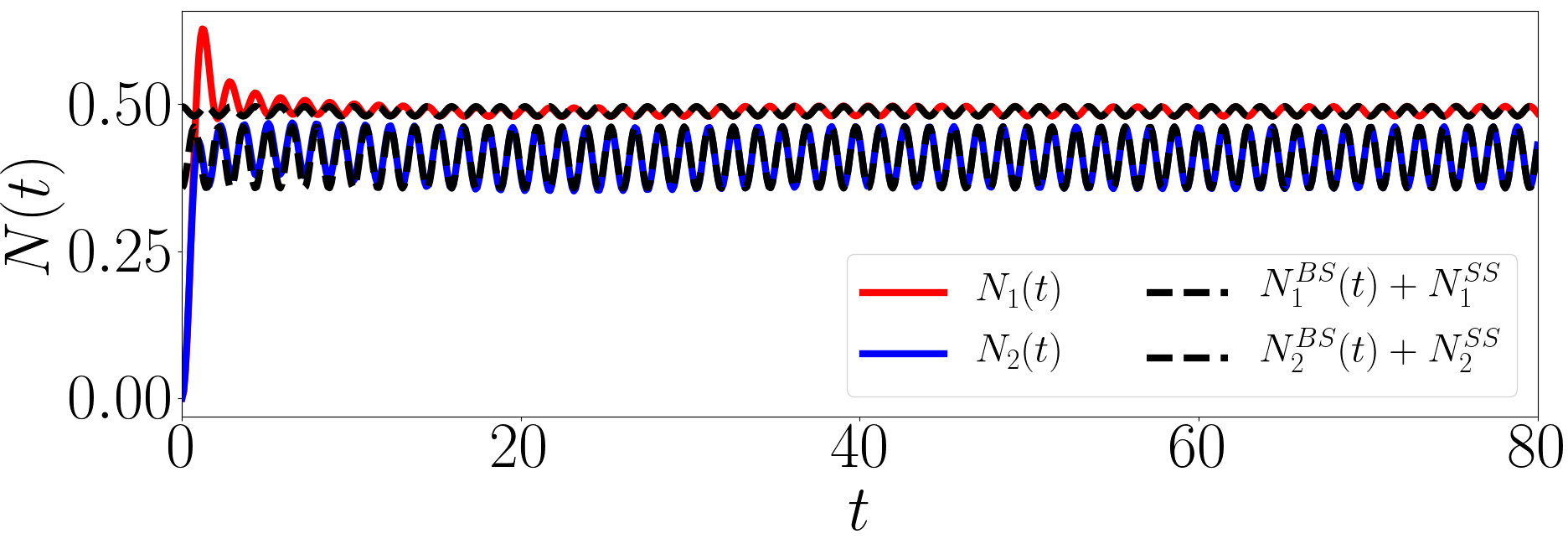}
	}	
	 \caption{(a) The variation of $\sum_{ij}\abs{I_{ij}}^2$ over different parameters of the Hamiltonian in Eq.~\eqref{nnh} with $\eta_s=V_L=V_R=1$ and $N=2$. Physically this quantity should identically vanish. We see that this happens only for  for certain parameter regimes of the Hamiltonian. In (b) we  verify that the non-zero values are associated to the presence of high energy bound states in the spectrum of the full system seen in Fig.~\eqref{bounds}. This plot shows the gap between the bound state energy and the edge of the band, $E_{bound}$, for the same parameters as in (a). We see that the value of $\Delta$ at which the bound state appears is exactly  the same value where the corresponding curves in (a) start taking non-zero values. (c) and (d) demonstrate that the numerical simulation for the two-point correlators agrees with  the analytic results obtained by  adding the bound state contributions to the  steady-state values. While (c) shows that the commutation relations are satisfied when we add the bound state contribution (d) on the other hand depicts the persistent oscillations in the particle densities  due to the bound states. Parameter values for these two plots are $ N=2$ for (c) and $N=3$ for (d),
		$N_b=100$, $\mu_R=1$, $\beta_R=10$, $\mu_L=0$, $\beta_L=0$, $V_L=V_R=\eta_w=1$, $\mu_w=0$, $\eta_b=1$ and  $\Delta=0.80$.}
\label{Ivar}
\end{figure}

\begin{figure}[htb!]
	\centering
	\subfigure[]{
		\includegraphics[width=40mm,height=40mm]{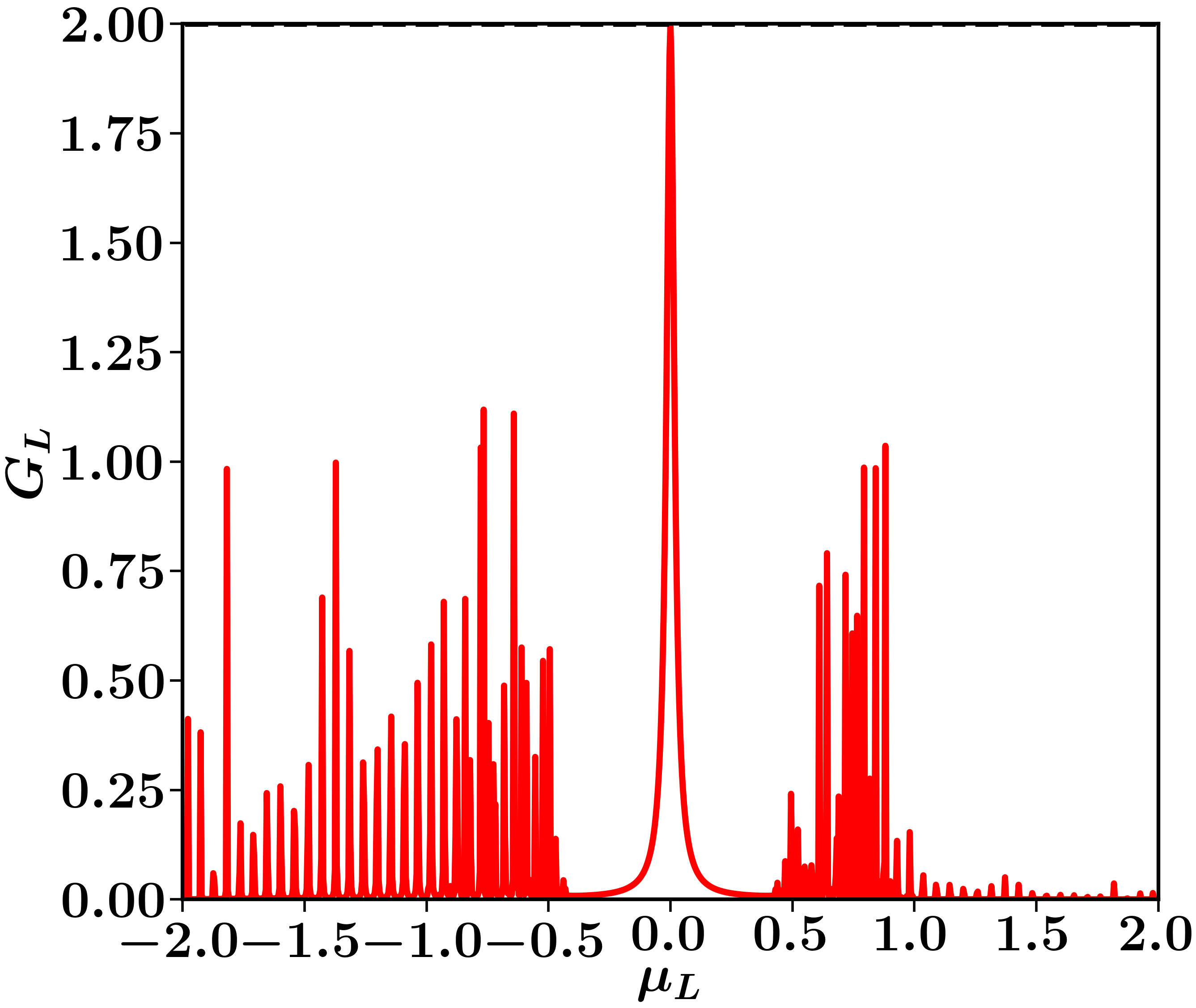}
	}
	\subfigure[]{
		\includegraphics[width=40mm,height=40mm]{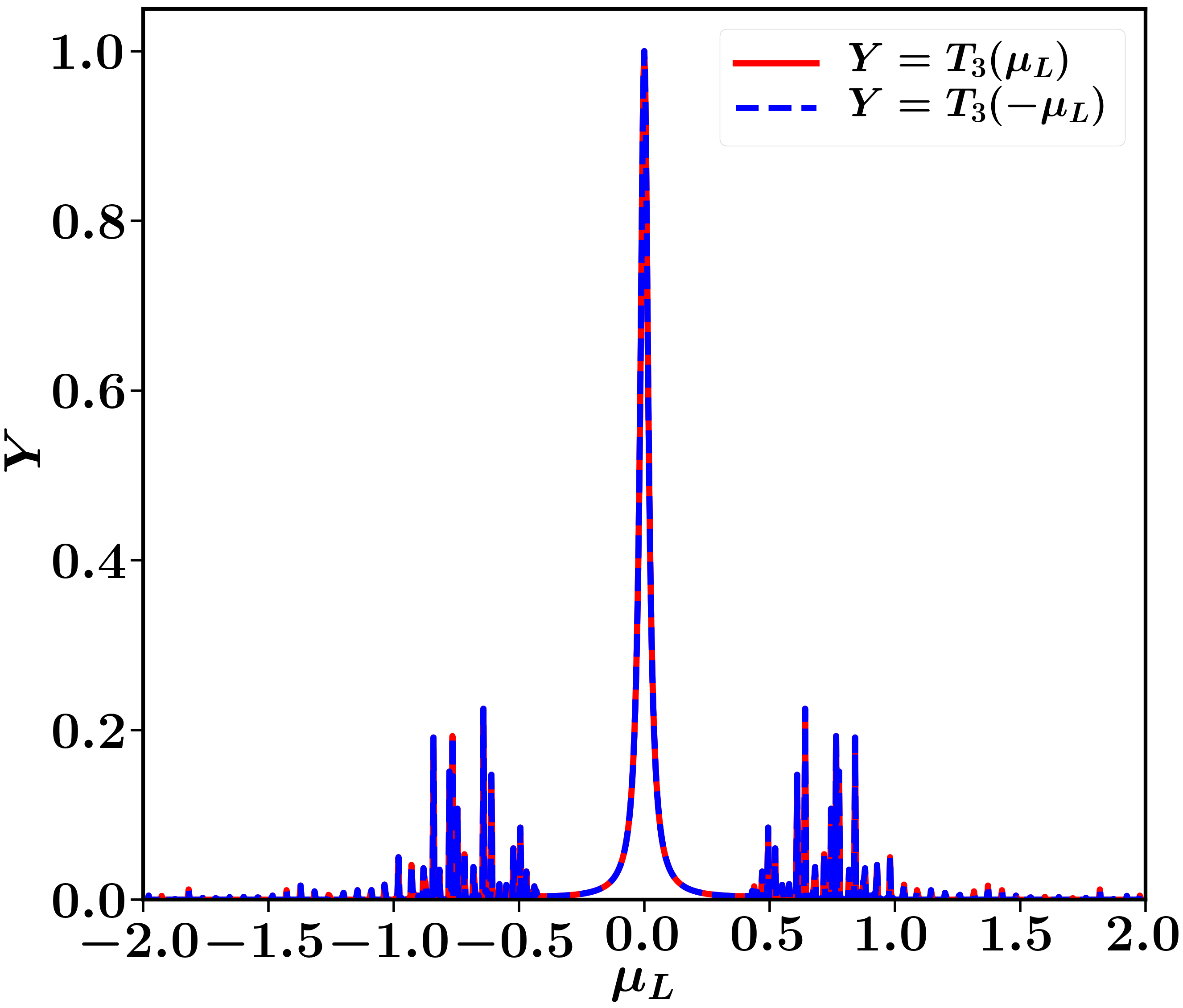}
	}
	\subfigure[]{
		\includegraphics[width=40mm,height=40mm]{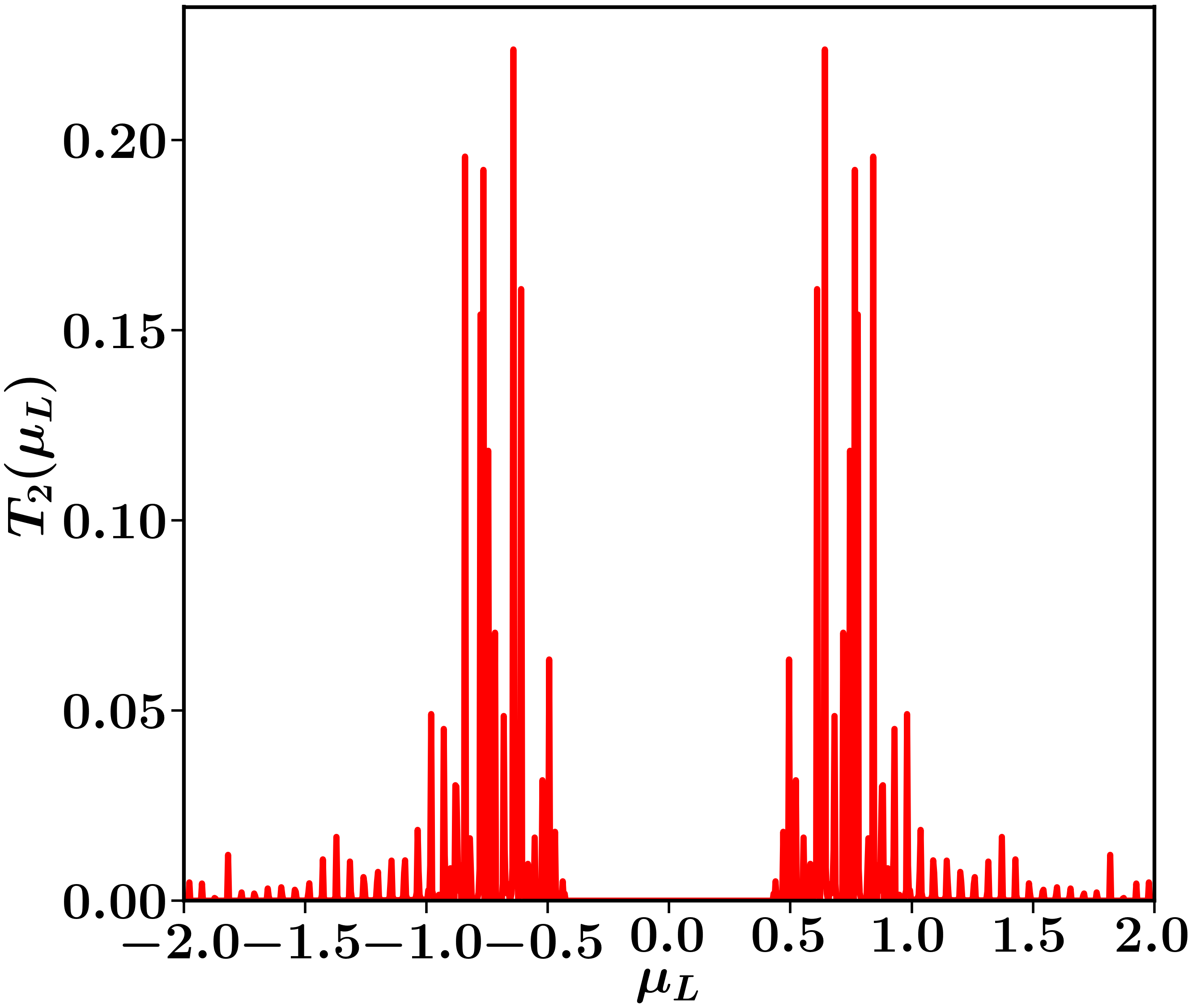}
	}
	\subfigure[]{
		\includegraphics[width=40mm,height=40mm]{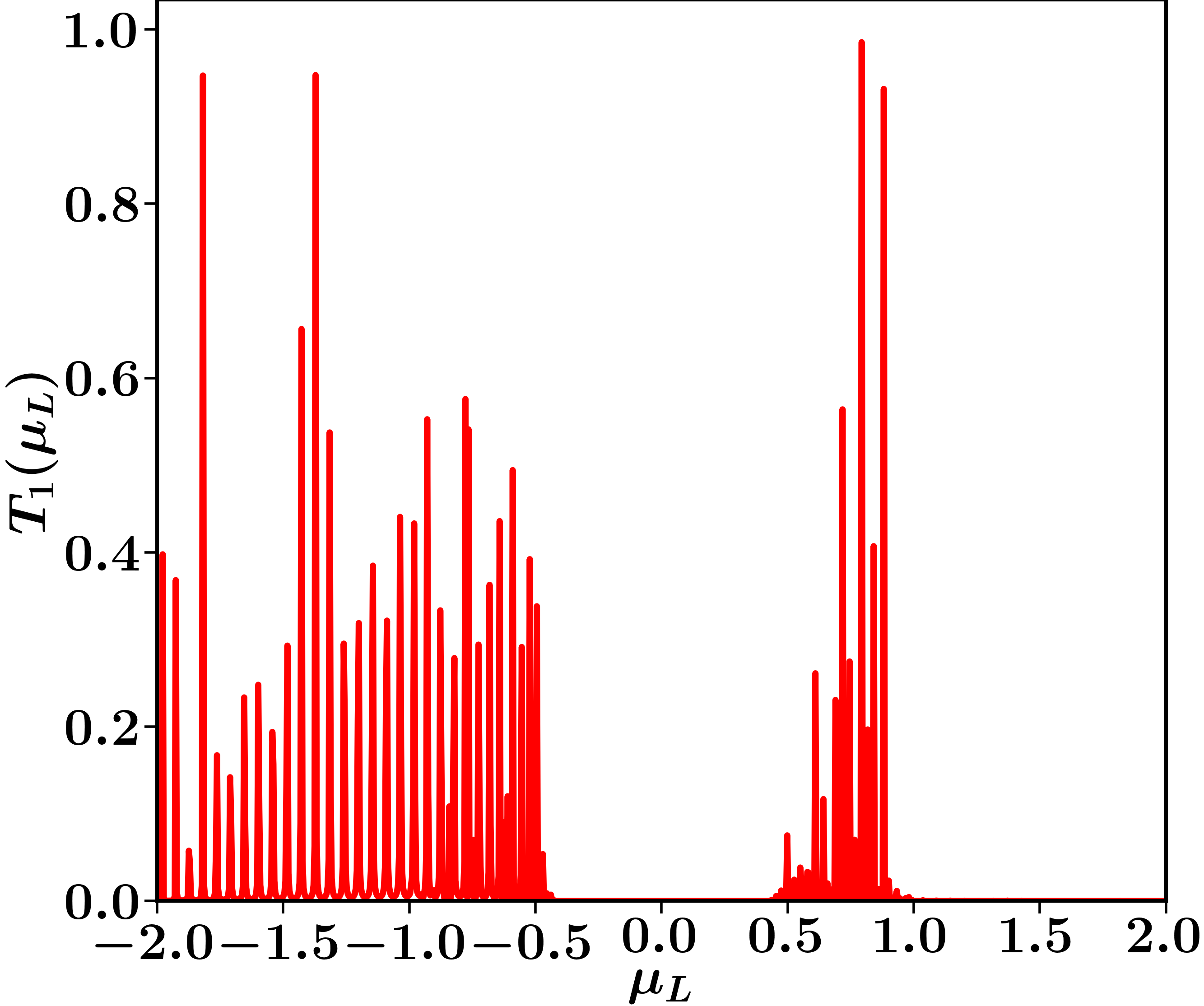}
	}
	\caption{Variation of zero temperature conductance and the terms contributing to it with the left bath chemical potential, $\mu_L$. Parameter values-- $ N=100$,
		$V_L=V_R=0.2$, $\eta_w=\mu_w=1$, $\eta_b=1$ and  $\Delta=0.25$. }
	\label{condplts}
\end{figure}
\begin{figure}[htb!]
	\centering
	\subfigure[]{
		\includegraphics[width=40mm,height=40mm]{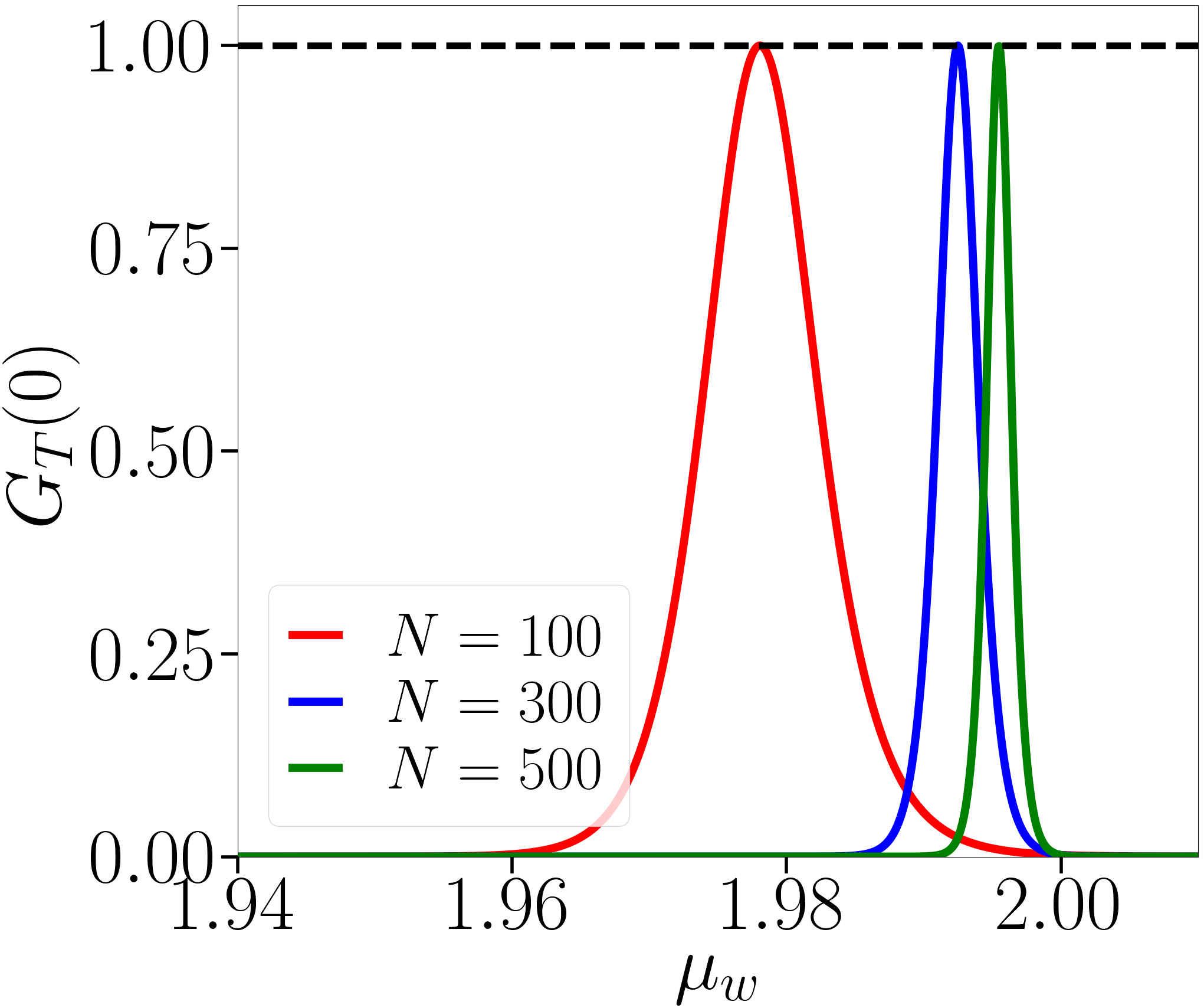}
	}
	\subfigure[]{
		\includegraphics[width=40mm,height=40mm]{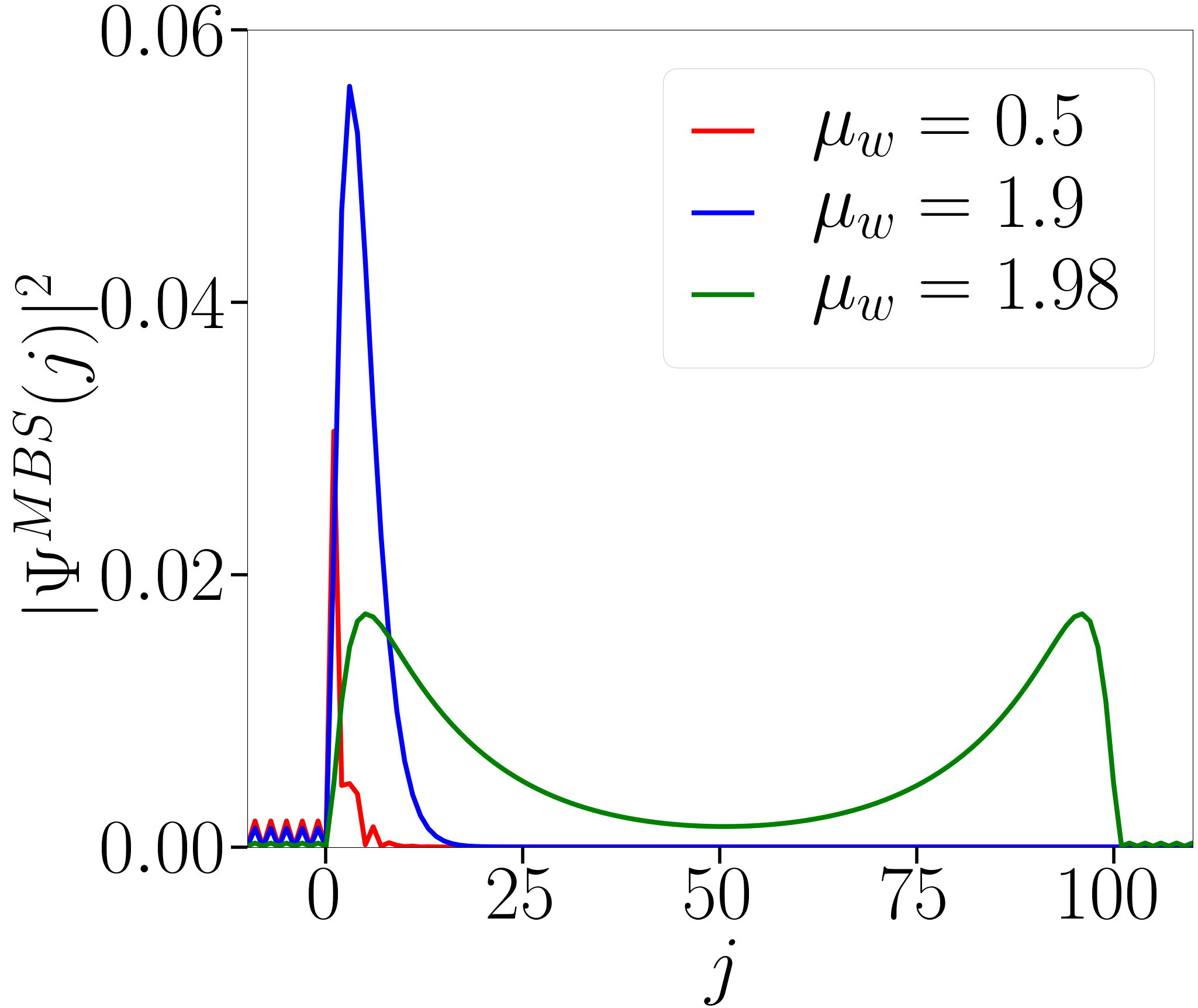}
	}
	\caption{(a)Variation of thermal conductance at $\mu_L=0$ in units of $\pi^2k_B^2T_L/6$ with the chemical potential, $\mu_w$, on the wire for different wire sizes. (b) shows the wave function of the  Majorana zero mode at $\mu_w$ far away and close to the topological phase transition point for $N=100$. The extended nature  of the Majorana wave functions spreading across the wire can be clearly seen when $\mu_w$ is close to the phase transition point. Other Parameter values-- $V_L=V_R=0.25$, $\eta_w=\eta_b=1$ and  $\Delta=0.3$. }
	\label{thcondplts}
\end{figure}

In general we find that, for the parameter regimes over which there exists a steady state, Eq.~\eqref{currexp} gives the value of the steady state current. We also verified that this expression reproduces the results given in Ref.~[\onlinecite{roy2012}]. As discussed in the end of Sec.~\eqref{sec:ness}, the non-vanishing of $I_{ij}$ in fact indicates  the presence of bound states which leads to the break-down of the NESS assumption. In Figs.~(\ref{bounds}a,\ref{bounds}b) we show the full energy spectrum of the system  for the parameter values  $ N=2$, $N_b=100$,  $V^L=V^R=\eta_s=1$, $\eta_b=2.5$ and for two values of  $\Delta$.  We see the appearance of a discrete energy level, indicating a bound state, for the parameter value  $\Delta=5$. 
In Fig.~(\ref{Ivar}a) we show the variation, with $\Delta$,  of the quantity  $\sum_{i,j} \abs{I_{ij}}^2$ for different parameter regimes of the Hamiltonian in Eq.~\eqref{nnh} with $N=2$. We see that the bound state contribution, for any fixed $\eta_b$, only kicks in after some critical value of $\Delta$. In Fig.~(\ref{Ivar}b) we show the variation of the energy gap, between the bound state level and the band edge, over the same parameter regimes as in Fig.~(\ref{Ivar}a). For any  fixed $\eta_b$, we see that the bound state appears at the same value of $\Delta$ as that where $I_{ij}$  in Fig.~(\ref{Ivar}a) becomes non-vanishing.  In Fig.~(\ref{Ivar}c) we plot the real and imaginary parts of the two point correlator $M_{12}=\langle a_1(t)a_2(t)\rangle$, in the presence of bound states, and compare the values obtained from the numerical simulations ($Y(t)$) with the analytic results [$M_{12}^{SS}+M_{12}^{BS}$, using Eqs.~(\ref{cccorr},\ref{Mlm})]. In the long time limit, the observed agreement requires that we add the bound state contributions   and we recover the correct anti-commutation relations. The bound state contribution leads to persistent oscillations in properties such as densities and this can be seen in   Fig.~(\ref{Ivar}d) where we compare the analytic result from Eq.~\ref{Nlm})  with the numerical simulation.

So far we have discussed short wires and highlighted the role of bound states and how the steady state description needs to be modified in their presence. We now briefly discuss the case of long Kitaev wires where we expect  topological phases with Majorana bound states. These zero-energy bound states lie inside the bath band widths and so do not lead to problems in the steady state. The conductance of the Kitaev chain is given by Eq.~\eqref{GLnegf1} and here we provide a numerical demonstration of the interpretation of the three transmission terms in this equation (see discussions after Eq.~\eqref{GLnegf1}).    
For a long wire, Fig.~(\ref{condplts}a) shows the conductance result which reveals the well known zero bias peak of strength $2$ in the topologically non trivial parameter regime,$|\mu_w|<2|\eta_w|$, of the wire Hamiltonian.  This peak is due to the existence of the zero energy Majorana bound states in this parameter regime. In Figs.~(\ref{condplts}b,\ref{condplts}c,\ref{condplts}d) we show the variation of the three terms which contribute to the conductance. As is clear from these plots, the peak is due to the perfect Andreev reflection at zero bias owing to the Majorana bound state. Within the superconducting gap, we also find $T_1(\mu_L)=T_2(\mu_L)=0$ for long wires which is consistent with their physical interpretation as normal and Andreev  transmission amplitudes. The transmission from left  to right bath for long wires(long enough so that the Majorana modes are isolated) can only occur via excitation of quasiparticles within the wire which is not possible if $\mu_L$ lies within the superconducting gap.   

 We now consider the thermal conductance  at the left end at zero bias $(\mu_L=0)$. This is given by Eq.~\eqref{econd} and we see that it is proportional to the net transmission, $T_1(0)+T_2(0)$. As we just discussed, if the wire is long enough so that the Majorana modes have no overlap within the wire, the net transmission and hence the zero bias thermal conductance would be zero. However, for a fixed finite size of the wire as we move closer to the topological phase transition point, $\mu_w=2\eta_w$, the spread of the Majorana zero modes in the wire increases and as we approach the phase transition point, these  modes hybridize to form two extended modes. This can be seen in Fig~.(\ref{thcondplts}b) where we plot the MBS wavefunctions for a few different $\mu_w$ values. This would then result in non-zero transmission probabilities at zero bias. As a result the thermal conductance for a finite sized wire will take finite values sufficiently close to the transition point. In Fig.~(\ref{thcondplts}a) we show the variation of thermal conductance at zero bias with $\mu_w$ while keeping other parameters fixed. We see that for a fixed $N$ the thermal conductance  peaks to a  strength $1$ (in units of the thermal conductance) at a $\mu_w$ value close to the infinite size phase transition point. Note that this means that  Andreev reflection is now suppressed and so the electrical conductance peak at zero bias would reduce from the value 2. We also see that with increasing system size, the the peak becomes narrower and moves closer  to the phase transition point. This peak has been noted recently~\cite{bondyopadhaya2020nonequilibrium} and we point out here that there are strong finite size effects and in particular the vanishing width with increasing wire size would make this difficult to observe experimentally.
  
\section{Next Nearest Neighbour Kitaev Chain}
\label{sec:long}

\begin{figure}[htb!]
	\centering
	\subfigure[$~\theta=0$]{
		\includegraphics[width=40mm,height=40mm]{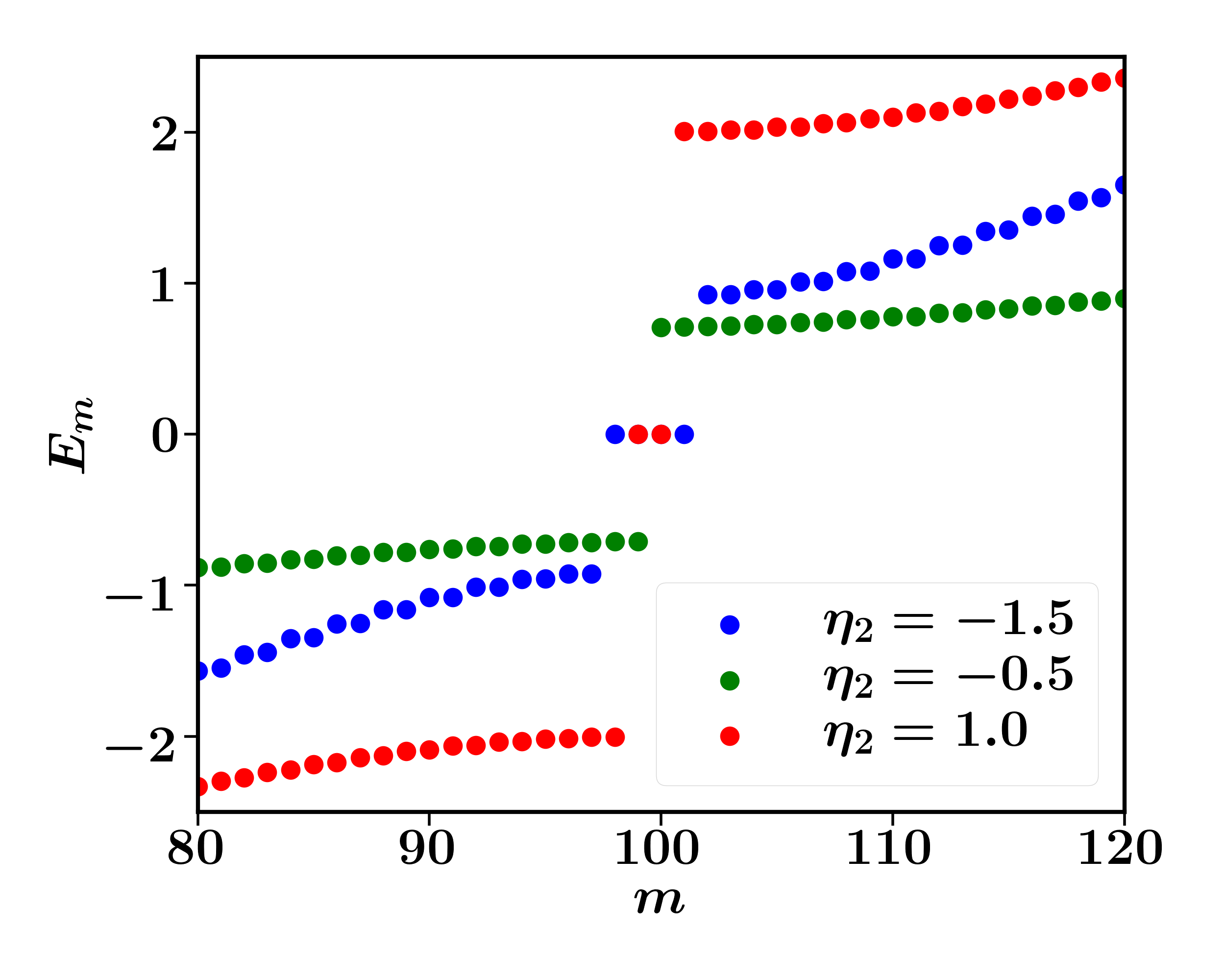}
}
	\subfigure[$~\theta=0.1$]{
		\includegraphics[width=40mm,height=40mm]{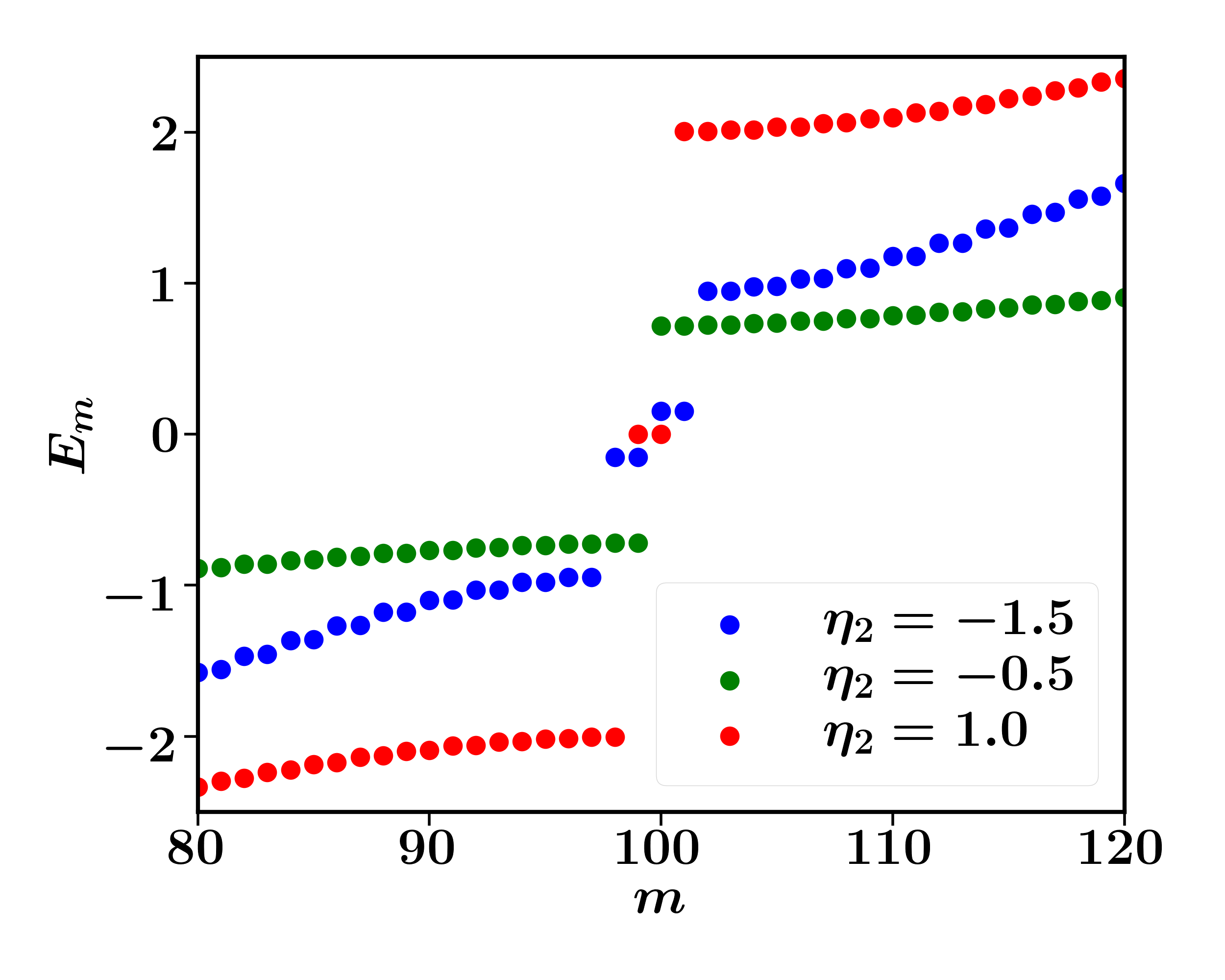}
	}
	\subfigure[$~\theta=0$]{
		\includegraphics[width=40mm,height=40mm]{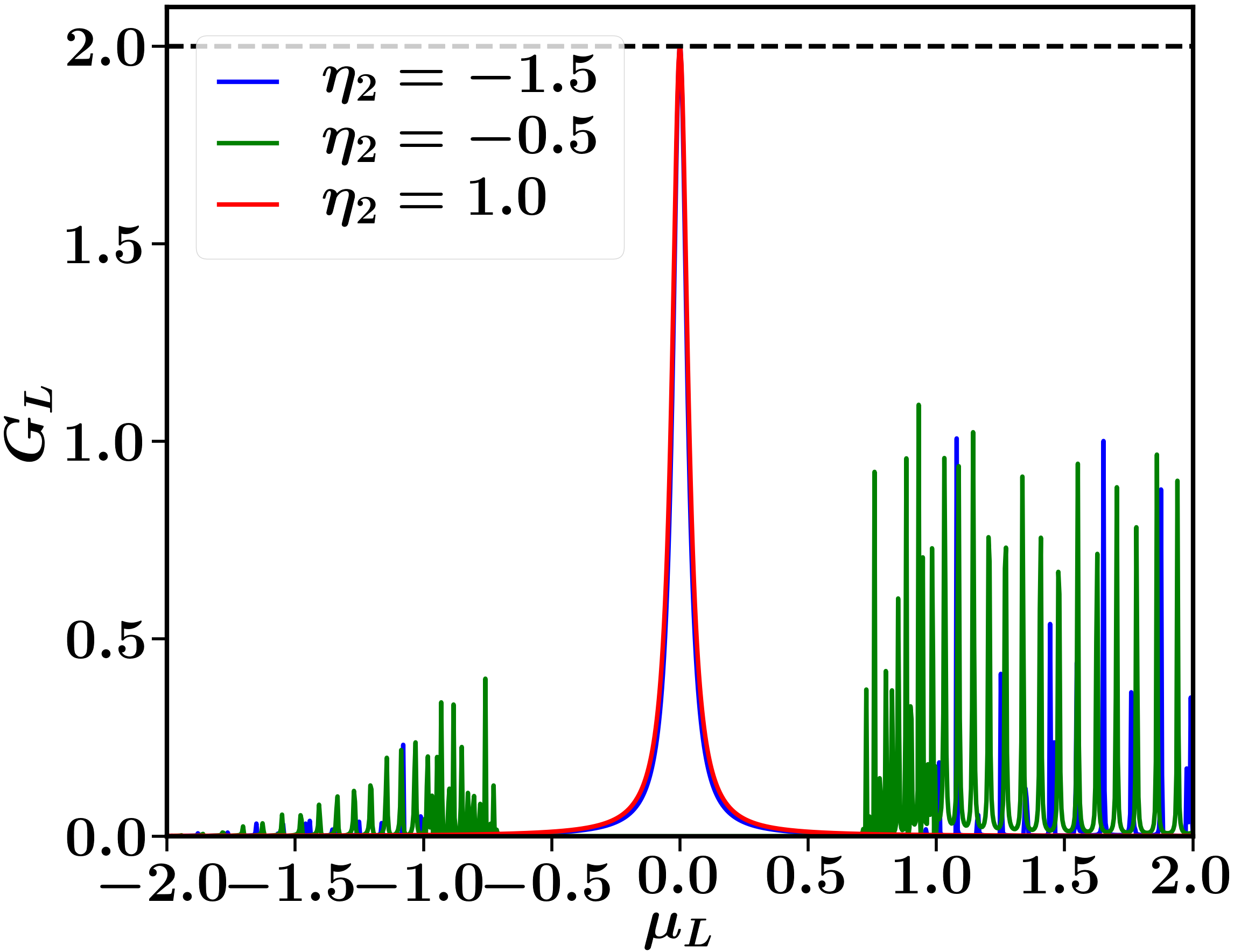}
	}
	\subfigure[$~\theta=0.1$]{
		\includegraphics[width=40mm,height=40mm]{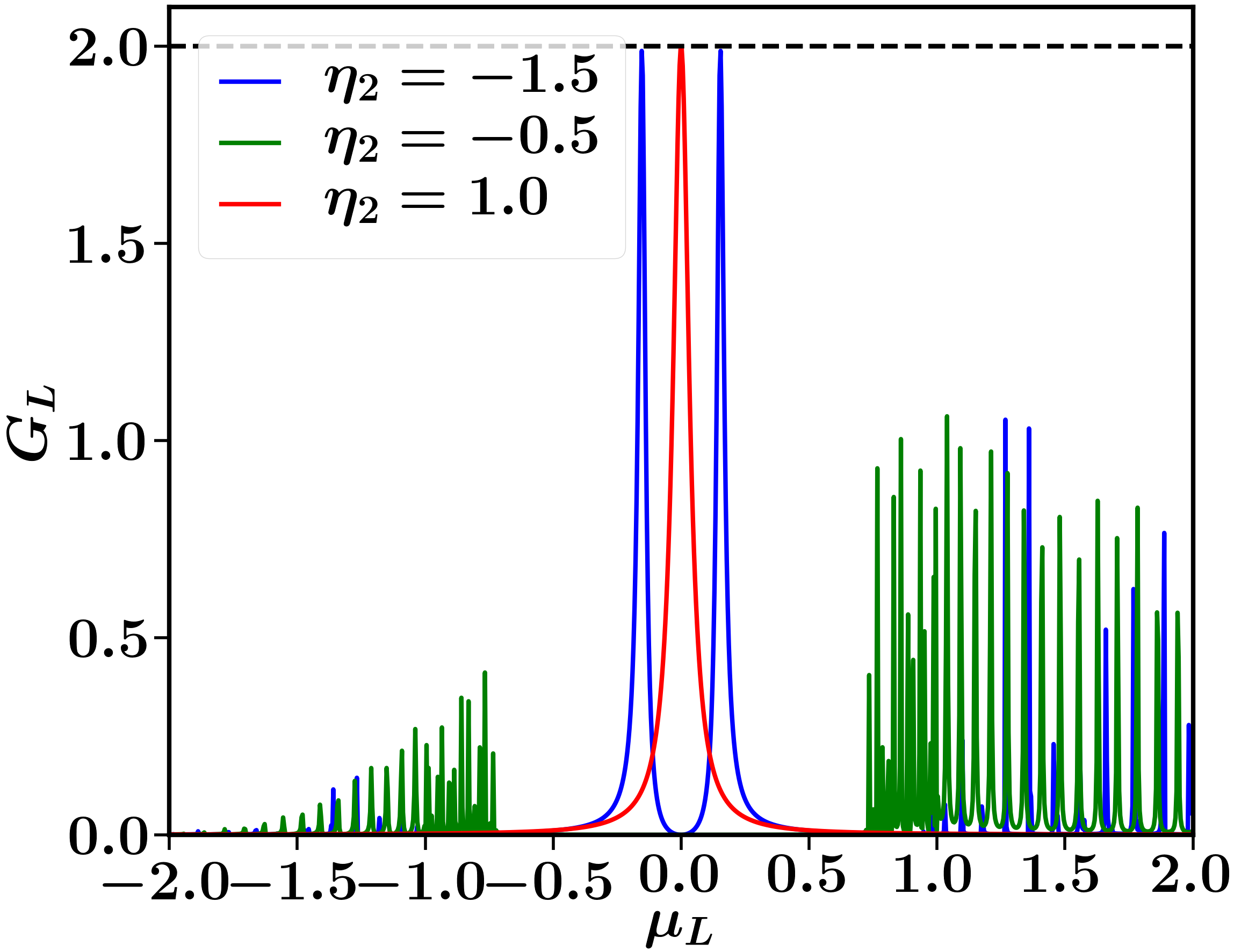}
	}
	\caption{(a) and (b)  show the spectrum of an isolated  next nearest neighbour Kitaev chain  at the  parameter values  $\eta_1=1$, $\mu_w=-2$, $N=100$, $V_L=V_R=0.3$ and $\eta_b=1.5$ at  zero and non-zero values of $\theta$ and for 
the values of $\eta_2=-1.5,~-0.5$ and $1$  corresponding to phase-2, phase-0 and phase-1 respectively.  The topological edge modes can be seen in the spectral plots in between the superconducting gap. The corresponding conductance results obtained from Eq.~\ref{GLnegf1} are shown in (c) and (d).  The conductance plots  for $\theta=0$ show a single zero bias peak for the two topologically non-trivial phases while  they show two distinct peaks for a non-zero $\theta$ value for phase-2.}
	\label{nnnplts}
\end{figure}

Finally in this section, as an application of our very general formalism, we discuss the transport properties of an open Kitaev chain with  interaction couplings that extend beyond nearest neighbours. In particular we  consider a wire Hamiltonian with  next nearest neighbour couplings:
\begin{align}
 \notag \mathcal{H}^W=\sum_{j=1}^N&\left[-\mu_wa_j^\dagger a_j-\eta_1\big( a_j^\dagger a_{j+1}+ a_ja_{j+1}+\text{c.c.}\big)\right.\\&-\eta_2\left.\big( a_j^\dagger a_{j+2}+ e^{i\theta} a_ja_{j+2}+\text{c.c.}\big)\right]. \label{NNNham}
\end{align}
The other parts of the total Hamiltonian (bath and wire-bath coupling) are taken to be the same as the Hamiltonian in Eq.~\ref{nnh}.
 Such models have  been discussed for the closed system and follow from Jordan Wigner transformation of a 1-D transverse Ising model with three spin interactions. The phase diagram of such a model reveals very interesting features~\cite{niu2012majorana}. For  $\theta=0$, one has two different topologically non-trivial phases, phase-1 and phase-2, apart from the topologically trivial phase (phase-0). Phase-1  and phase-2 contain one and two zero modes localized at each end respectively. However, for $\theta\neq 0$ the energy of the localized modes of phase-2  is lifted from zero to a finite value. These facts can be seen in the spectrum of the isolated wire Hamiltonian in the three phases for the case of zero and non-zero  $\theta$ and we show this in Fig.~(\ref{nnnplts}a) and Fig.~(\ref{nnnplts}b). The different values of $\eta_2=-1.5,0.5$, and $1$ in these spectral plots  lie in the parameter regimes of phase-2, phase-0 and phase-1 respectively. The topological edge modes can be seen between the superconducting gap near zero energy.
 
In  Fig.~(\ref{nnnplts}c) and  Fig.~(\ref{nnnplts}d) we show the conductance results of this model obtained from Eq.~\ref{GLnegf1}, evaluated  with the Hamiltonian in Eq.~\eqref{NNNham}, for the parameter regimes of the three phases at zero and non-zero $\theta$ respectively. We see that for $\theta=0$ there is a single zero bias peak as all topological edge modes in the three phases are of zero energy. However, for $\theta\neq 0$ we see two distinct peaks of strength $2$ for the case of phase-2 at the same energy as the energy of the topological edge mode. Thus  phase-2 is accompanied by two perfect Andreev reflections at energies of the edge modes and the degeneracy of the modes localized at the two ends is broken. Phase-1 on the other hand is has a single peak at zero energy for $\theta$ being zero or non-zero representing the fact that the degeneracy of the zero modes at the two ends is not lifted. Phase-0 as usual has no such peak  since there are no edge modes. The splitting of the conductance peak seen in  Fig.~(\ref{nnnplts}d) has recently been observed in a ladder system of two coupled chains with longe range interactions (power-law form) within each chain \cite{nehra2020}.

\section{Discussion}
\label{sec:summary}
In conclusion, we considered transport in a wire that is modelled by a spinless superconductor with a mean-field pairing form for the interaction term, so that it is effectively described by a general quadratic Hamiltonian. We investigated transport in the wire for the so-called N-S-N geometry where the superconductor is placed between normal leads. Thus, in our set-up, the wire is attached to free electron baths at different temperatures and chemical potentials and we investigated particle and energy transport,  using the open system framework of quantum Langevin equations (QLE) and nonequilibrium Green's function (NEGF). 

Our main results are the exact analytic expressions for the  particle current, energy current, other two-point correlations in the nonequilibrium steady state and high energy bound state corrections to them. These have the same structure as NEGF expressions for free electrons, but now involve two sets of Green's functions. We show that  current expressions generically involve  three types of terms  which can be physically interpreted in terms of normal and Andreev processes. We also derive the Landauer formula for the heat current and show that the Andreev reflection process does not contribute to this leading to absence of a zero bias peak in the thermal conductance for parameter regimes far away from the topological phase transition point. However, for fixed wire sizes the zero bias thermal conductance shows a peak as one moves sufficiently close the transition point.
To derive these expressions one has to assume the existence of a nonequilibrium steady state and we  relate this to the existence of bound states (discrete energy levels) in the spectrum of the entire coupled system of the superconducting wire and the  baths. The role of bound states on the existence of steady states is known for normal systems~\cite{dhar2006,stefanucci2007,khosravi2008,jussiau2019} and has recently been investigated numerically for the case of superconductors in Ref.~[\onlinecite{roy2019}]. In the present work, we examined this issue analytically for a general spinless superconductor and obtained explicit expressions for the bound state contributions to the two point correlators.

Next, by performing  an exact numerical diagonalization of the full quadratic Hamiltonian of wire and baths, we computed the time evolution of the current and local densities, starting from the same initial density matrix as used in the steady state calculations. We then showed that the results from this approach agree perfectly with the analytic expressions, both for the case where there are no high energy bound states and also in the presence of such states after we add the additional correction terms to the usual steady state results. We verify from the numerics that these correction terms are crucial in ensuring that the fermionic anti-commutation relations are satisfied. Our analytic results for the bound states also  reproduce the persistent temporal oscillations observed in the steady state.

 Finally, as an application of our general formalism, we investigated the conductance results of long wires for a Kitaev chain with next neighbour interactions. In particular it is known that several interesting topological phases are obtained if one adds a phase difference between the superconducting parings of the nearest neighbour and next neighbour terms.  We found that the conductance peak of strength $2$ exists whenever the system is in a topologically non-trivial phase.  However, in the topologically non-trivial phase-3 the presence of the phase difference leads to a lifting of the degeneracy of the topological edge modes and consequently the conductance results showed two distinct peaks at the new energies of the edge modes. Our general formalism can be readily applied to other physically interesting set-ups such as those studied in Refs.~\onlinecite{wakatsuki2014,vodola2015long}. 

In the present work we have not imposed the self-consistency condition for the superconducting pairing terms and  we saw that this leads to the non-conservation of particle number in the wire. Physically this corresponds to the situation of proximity-induced superconductivity, where the wire is placed on a superconducting substrate that is grounded and hence serves as a sink (or source) of electrons. Another interesting situation  would be one where one is  actually looking at transport through a superconductor and the self-consistency condition has to be introduced. This appears to be quite non-trivial and our results, which provide analytic expressions for all two-point correlations in the nonequilibrium steady state, provides a good starting point to study this problem. 

\begin{acknowledgments}
  We thank Subhro Bhattacharjee for useful discussions. J.M.B. thanks Abhishodh Prakash and Adhip Agarwala for their help in doing this work.  AD thanks Dibyendu Roy, Diptiman Sen and Sankar Das Sarma for useful discussions. 
We acknowledge support of the Department of Atomic Energy, Government of India, under project no.12-R$\&$D-TFR-5.10-1100.
\end{acknowledgments}

\bibliography{biblio}
\appendix
\section{Derivation of the current expression}
 We present the derivation of the expression for the current  in Eq.~\eqref{currexp} here. We start by substituting Eq.~\eqref{ftb} in Eq.~\eqref{currfr1} we get,

\begin{align}
J_L=2\notag\text{Im}\Big\lbrace\int &d\omega d\omega^\prime e^{i(\omega-\omega^\prime)t}\\&\expval{c_m^\dagger(\omega)\lbrace\eta_m^L(\omega^\prime)+[\Sigma_L^+(\omega^\prime)]_{ml}\tilde{c}_l(\omega^\prime)\rbrace}\Big\rbrace
\end{align}

Using Eq.~\eqref{ngef} in the above expression we have,

\begin{align}
\notag\Big \langle c_m^\dagger(\omega)\lbrace\eta_m^L(\omega^\prime)&+[\Sigma_L^+(\omega^\prime)]_{ml}\tilde{c}_l(\omega^\prime)\rbrace\Big\rangle\\&= A_1 + A_2+ A_3+A_4 +A_5\label{a1}
\end{align}

where, 

\begin{widetext}
\begin{eqnarray}
\notag A_1&&=\int_{-\infty}^\infty \int_{-\infty}^\infty d\omega d\omega^\prime e^{i(\omega-\omega^\prime)t}[G_1^-(\omega)]_{km}\expval{\eta_k^{L\dagger}(\omega)\eta_m^{L}(\omega^\prime)}=\int_{-\infty}^\infty d\omega \Tr[G_1^-(\omega)\Gamma_L(\omega)]f_L(\omega)
\end{eqnarray}
\begin{eqnarray}
\notag	A_2&&=\int_{-\infty}^\infty \int_{-\infty}^\infty d\omega d\omega^\prime e^{i(\omega-\omega^\prime)t}[G_1^-(\omega)]_{km}[\Sigma_L^+(\omega^\prime)]_{ml}[G_1^+(\omega^\prime)]_{lk^\prime}\notag\expval{\eta_k^{L\dagger}(\omega)\eta_{k^\prime}^{L}(\omega^\prime)}\\&&=\int_{-\infty}^{\infty}d\omega
\Tr[G_1^-(\omega)\Sigma_L^+(\omega)G_1^+(\omega)\Gamma_L(\omega)]f_L(\omega)\\
\notag	A_3&&=\int_{-\infty}^\infty \int_{-\infty}^\infty d\omega d\omega^\prime e^{i(\omega-\omega^\prime)t} [G_1^-(\omega)]_{km}[\Sigma_L^+(\omega^\prime)]_{ml}[G_1^+(\omega^\prime)]_{lk^\prime}\notag\expval{\eta_k^{R\dagger}(\omega)\eta_{k^\prime}^{R}(\omega^\prime)}\\&&=\int_{-\infty}^{\infty}d\omega \Tr[G_1^-(\omega)\Sigma_L^+(\omega)G_1^+(\omega)\Gamma_R(\omega)]f_R(\omega)\\
\notag	A_4&&=\int_{-\infty}^\infty \int_{-\infty}^\infty d\omega d\omega^\prime e^{i(\omega-\omega^\prime)t} [G_2^-(\omega)]_{km}[\Sigma_L^+(\omega^\prime)]_{ml}[G_2^+(\omega^\prime)]_{lk^\prime}\notag\expval{\eta_k^{L}(-\omega)\eta_{k^\prime}^{L^\dagger}(-\omega^\prime)}\\&&=\int_{-\infty}^{\infty}d\omega \Tr[G_2^-(\omega)\Sigma_L^+(\omega)G_2^+(\omega)\Gamma_L^T(-\omega)](1-f_L(-\omega))\\
\notag    A_5&&=\int_{-\infty}^\infty \int_{-\infty}^\infty d\omega d\omega^\prime e^{i(\omega-\omega^\prime)t} [G_2^-(\omega)]_{km}[\Sigma_L^+(\omega^\prime)]_{ml}[G_2^+(\omega^\prime)]_{lk^\prime}\expval{\eta_k^{R}(-\omega)\eta_{k^\prime}^{R\dagger}(-\omega^\prime)}\notag\\&&=\int_{-\infty}^{\infty}d\omega \Tr[G_2^-(\omega)\Sigma_L^+(\omega)G_2^+(\omega)\Gamma_R^T(-\omega)](1-f_R(-\omega))    	
\end{eqnarray}
\end{widetext}

The imaginary parts of $A_1$, $A_2$, $A_3$, $A_4$ and $A_5$ can be shown to be the following, 
\begin{eqnarray}
&&\notag\Im{A_1}=
\\&&\int_{-\infty}^\infty d\omega \Tr[\frac{G_1^-(\omega)-G_1^+(\omega)}{2i}\Gamma_L(\omega)]f_L(\omega)\label{imt1}\\
&&\frac{-1}{\pi}\Im{A_2}=\\&&\int_{-\infty}^{\infty}d\omega\notag \Tr[G_1^-(\omega)\Gamma_L(\omega)G_1^+(\omega)\Gamma_L(\omega)]f_L(\omega)\notag\\
&&\frac{-1}{\pi}\Im{A_3}=\\&&\int_{-\infty}^{\infty}d\omega \notag \Tr[G_1^-(\omega)\Gamma_L(\omega)G_1^+(\omega)\Gamma_R(\omega)]f_R(\omega)\\
&&\frac{1}{\pi}\Im{A_4}=\\&&\int_{-\infty}^{\infty}d\omega\notag \Tr[G_2^-(\omega)\Gamma_L(\omega)G_2^+(\omega)\Gamma_L^T(-\omega)](f_L(-\omega)-1)\notag\\
&&\frac{1}{\pi}\Im{A_5}=\\&&\notag\int_{-\infty}^{\infty}d\omega \Tr[G_2^-(\omega)\Gamma_L(\omega)G_2^+(\omega)\Gamma_R^T(-\omega)](f_R(-\omega)-1)\notag
\end{eqnarray}
It is fairly straightforward to show that,
\begin{align}
\notag[G_1^-(\omega)-G_1^+(\omega)]/2i&=\pi \big[ G_1^+(\omega)(\Gamma_L(\omega)+\Gamma_R(\omega))G_1^-(\omega)\notag\\&+G_2^+(\omega)(\Gamma_L^T(-\omega)+\Gamma_R^T(-\omega))G_2^-(\omega)\big]\notag
\end{align}
\vspace{1mm}
Substituting this result in Eq.~\eqref{imt1} and adding up the imaginary parts of the terms $A_1$, $A_2$, $A_3$, $A_4$ and $A_5$, we obtain the required expression for the current entering the wire from the left reservoir to be
\begin{align}
\notag &\frac{J_L}{2\pi}=\\\notag&\int_{-\infty}^{\infty}d\omega \bigg(\Tr[G_1^+(\omega)\Gamma_R(\omega)G_1^-(\omega)\Gamma_L(\omega)](f_L^e(\omega)-f_R^e(\omega))\\\notag&+\Tr[G_2^+(\omega)\Gamma_R^T(-\omega)G_2^-(\omega)\Gamma_L(\omega)](f_L^e(\omega)-f_R^h(\omega))\\\notag&+\Tr[G_2^+(\omega)\Gamma_L^T(-\omega)G_2^-(\omega)\Gamma_L(\omega)](f_L^e(\omega)-f_L^h(\omega))\bigg)\notag
\end{align}
The current from the right reservoir into the wire, $J_R$ can be obtained with similar algebra and is given by,
\begin{align}
\notag &\frac{J_R}{2\pi}=\\\notag&\int_{-\infty}^{\infty}d\omega \bigg(\Tr[G_1^+(\omega)\Gamma_L(\omega)G_1^-(\omega)\Gamma_R(\omega)](f_R^e(\omega)-f_L^e(\omega))\\\notag&+\Tr[G_2^+(\omega)\Gamma_L^T(-\omega)G_2^-(\omega)\Gamma_R(\omega)](f_R^e(\omega)-f_L^h(\omega))\\\notag&+\Tr[G_2^+(\omega)\Gamma_R^T(-\omega)G_2^-(\omega)\Gamma_R(\omega)](f_R^e(\omega)-f_R^h(\omega))\bigg)\notag
\end{align}

\section{Derivation of the expressions for bound sate contribution to the correlators}
\label{appendix2}
To obtain the contribution of high energy bound states to the two point correlators we begin by considering form of the Hamiltonian written in Eq.~\ref{Zhamil}. Clearly, the equations of motion for the entire system could be written as, 
\begin{align}
\begin{pmatrix}
\dot C(t)\\\dot C^\dagger(t)
\end{pmatrix}=-i\mathcal{Z}\begin{pmatrix}
C(t)\\ C^\dagger(t)
\end{pmatrix}
\end{align}
where $C(t)$ and $\mathcal{Z}$ are the same as in section~\ref{sec:timeevol}. This directly gives us the full solution of to  be,
\begin{equation}
c_{l}(t)=i[\mathcal{G}_1(t)]_{lm} c_m+i [\mathcal{G}_2(t)]_{lm} c^\dagger_{m} 
\end{equation}
where,
\begin{equation}
\mathcal{G}(t)=-ie^{-i\mathcal{Z}t}\theta(t)=\begin{pmatrix}\mathcal{G}_1(t) && \mathcal{G}_2(t)\\\mathcal{G}_2^*(t) && \mathcal{G}_1^*(t)\end{pmatrix}.
\end{equation}
We can also expand this Green's function in terms of the eigen vectors of the matrix $\mathcal{Z}$ as,
\begin{align}
	[\mathcal{G}_1(t)]_{pq}&=-i\sum_{E} e^{-iEt}\Psi_E(p)\Psi_E^*(q)~~\text{and}\\
	[\mathcal{G}_2(t)]_{pq}&=-i\sum_{E} e^{-iEt}\Psi_E(p)\Phi_E^*(q).
\end{align}
where $\begin{pmatrix}\Psi_E \\ \Phi_E \end{pmatrix}$ is the eigenvector of the matrix $\mathcal{Z}$ with energy $E$. In the long time limit, it can be shown that these expressions reduce to~\cite{dhar2006},
\begin{align}
\lim_{t\rightarrow\infty}[\mathcal{G}_1(t)]_{pq}&=-i\sum_{E_b} e^{-iE_bt}\Psi_{E_b}(p)\Psi_{E_b}^*(q)~~\text{and}\\
\lim_{t\rightarrow\infty}[\mathcal{G}_2(t)]_{pq}&=-i\sum_{E_b} e^{-iE_bt}\Psi_{E_b}(p)\Phi_{E_b}^*(q).
\end{align}
The sum now runs only over the bound states of the Hamiltonian.
The Fourier transform of this green's function is given by,
\begin{equation}
	\tilde{\mathcal{G}}(\omega)=\begin{pmatrix}\tilde{\mathcal{G}}_1(\omega) && \tilde{\mathcal{G}}_2(\omega)\\\tilde{\mathcal{G}}_2^*(\omega) && \tilde{\mathcal{G}}_1^*(\omega)\end{pmatrix}=\frac{1}{\omega+i\epsilon-\mathcal{Z}}\label{tildeMGomega}
\end{equation}
Using these Green's function we can obtain the two point correlators of the system as a sum of the steady state contribution and the bound state contribution. To see this explicitly, we first need to relate the components Green's functions $\tilde{\mathcal{G}}_1(\omega)$ and $\tilde{\mathcal{G}}_2(\omega)$ with the Green's functions $G_1^+(\omega)$ and $G_2^+(\omega)$. Note that $G_1^+(\omega)$ and $G_2^+(\omega)$ are matrices of dimension $N$ while $\tilde{\mathcal{G}}_1(\omega)$ and $\tilde{\mathcal{G}}_2(\omega)$ are matrices of dimension $N_S$. So, we split  $\tilde{\mathcal{G}}_1(\omega)$ and $\tilde{\mathcal{G}}_2(\omega)$ as follows,
\begin{equation}
	\tilde{\mathcal{G}}_1(\omega)=\begin{pmatrix}
	G_1^W(\omega)&& G_1^{WL}(\omega) && G_1^{WR}(\omega) \\ G_1^{LW}(\omega) && G_1^{L}(\omega) && G_1^{LR}(\omega) \\ G_1^{RW}(\omega)&& G_1^{RL}(\omega) && G_1^{R}(\omega) 
	\end{pmatrix}
\end{equation}
where the components of  $G_1^W(\omega)$ are given by $[\tilde{\mathcal{G}}_1(\omega)]_{ij}$, the components of  $G_1^{WL}(\omega)$ are given by $[\tilde{\mathcal{G}}_1(\omega)]_{i\alpha}$ and like wise for the other  matrices in this equation. We remind the reader here that $i,j,m,n$ denote the sites on the wire while as $\alpha,\mu,\nu$ and $\alpha',\mu',\nu'$ denote left bath and right bath sites respectively.  $\tilde{\mathcal{G}}_2(\omega)$ can be split similarly.  We now rewrite Eq.~\ref{tildeMGomega} in the following block form,
\begin{widetext}
\begin{equation}	\begingroup 
\setlength\arraycolsep{1pt}
\notag\begin{pmatrix}
\omega-H_W && -V_L && -V_R && -K && 0 && 0 \\ -V_L^\dagger && \omega-H_L&& 0 && 0 && 0 && 0\\ -V_R^\dagger && 0 &&\omega-H_R&&0&&0&&0\\K^*&& 0&& 0&&\omega+H_W^*&& V_L^* && V_R^*\\ 0&&0&& 0&& V_L^{\dagger*} &&\omega+ H_L^*&& 0\\ 0 && 0 && 0&&V_R^{\dagger*} && 0 && \omega+H_R^*
\end{pmatrix}\\	\begin{pmatrix}
G_1^W&& G_1^{WL} && G_1^{WR}&& G_2^W && G_2^{WL} && G_2^{WR} \\ G_1^{LW} && G_1^{L} && G_1^{LR} && G_2^{LW} && G_2^{L} && G_2^{LR}\\ G_1^{RW}&& G_1^{RL} && G_1^{R} && G_2^{RW}&& G_2^{RL} && G_2^{R}\\ G_2^{W*} && G_2^{WL*} && G_2^{WR*}&& G_1^{W*} && G_1^{WL*} && G_1^{WR*}\\G_2^{LW*} && G_2^{L*} && G_2^{LR*}&& G_1^{LW*} && G_1^{L*} && G_1^{LR*}\\G_2^{RW*} && G_2^{RL*} && G_2^{R*}&& G_1^{RW*} && G_1^{RL*} && G_1^{R*}
\end{pmatrix}=I	
\endgroup
\end{equation}	
\end{widetext}
From this we obtain the following required relations,
\begin{align}
&G_1^W(\omega)=G_1^+(\omega)\label{G1W}~~,~~G_2^W(\omega)=G_2^+(-\omega)\\
&G_1^{WL}(\omega)=G_1^{+}(\omega)V_L g_L^+(\omega)\\
&G_1^{WR}(\omega)=G_1^{+}(\omega)V_R g_R^+(\omega)
\end{align}
\begin{align}
&G_2^{WL}(\omega)=G_2^{+}(-\omega)V_L^* g_L^{+*}(\omega)\\
&G_2^{WR}(\omega)=G_2^{+}(-\omega)V_R^* g_R^{+*}(\omega)\label{G2WR}
\end{align}
 We can now consider the two point correlators  of the wire operators. Assuming initially that there is no correlation between the wire and the baths we  can be write these as,
 \begin{widetext}
\begin{align}
\expval{c_l^\dagger(t)c_m(t)}\notag&=\sum_{\alpha\nu}\bigg[[\mathcal{G}_1^\dagger(t)]_{\alpha l}\expval{c_\alpha^\dagger c_\nu}[\mathcal{G}_1(t)]_{m \nu}+[\mathcal{G}_2^\dagger(t)]_{\alpha l}\expval{c_\alpha c_\nu ^\dagger}[\mathcal{G}_2(t)]_{m \nu}\bigg]
\\&+\sum_{\alpha^\prime\nu^\prime}\bigg[[\mathcal{G}_1^\dagger(t)]_{\alpha^\prime l}\expval{c_{\alpha^\prime}^\dagger c_{\nu^\prime}}[\mathcal{G}_1(t)]_{m \nu^\prime}+[\mathcal{G}_2^\dagger(t)]_{\alpha^\prime l}\expval{c_{\alpha^\prime} c_{\nu^\prime} ^\dagger}[\mathcal{G}_2(t)]_{m \nu^\prime}\bigg]
\\&\notag+\sum_{ij}\bigg[[\mathcal{G}_1^\dagger(t)]_{i l}\expval{c_{i}^\dagger c_{j}}[\mathcal{G}_1(t)]_{m j}+[\mathcal{G}_2^\dagger(t)]_{i l}\expval{c_{i} c_{j} ^\dagger}[\mathcal{G}_2(t)]_{m j}\bigg]
\end{align}
\begin{align}
\expval{c_l(t)c_m(t)}\notag&=\sum_{\alpha\nu}\bigg[[\mathcal{G}_1(t)]_{l \alpha }\expval{c_\alpha c^\dagger_\nu}[\mathcal{G}_2(t)]_{m \nu}+[\mathcal{G}_2(t)]_{l \alpha}\expval{c_\alpha^\dagger c_\nu}[\mathcal{G}_1(t)]_{m \nu}\bigg]
\\&+\sum_{\alpha^\prime\nu^\prime}\bigg[[\mathcal{G}_1(t)]_{l \alpha^\prime}\expval{c_{\alpha^\prime} c_{\nu^\prime}^\dagger}[\mathcal{G}_2(t)]_{m \nu^\prime}+[\mathcal{G}_2(t)]_{l \alpha^\prime}\expval{c_{\alpha^\prime}^\dagger c_{\nu^\prime}}[\mathcal{G}_1(t)]_{m \nu^\prime}\bigg]
\\&\notag+\sum_{ij}\bigg[[\mathcal{G}_2(t)]_{ l i}\expval{c_{i}^\dagger c_{j}}[\mathcal{G}_1(t)]_{m j}+[\mathcal{G}_1(t)]_{ l i}\expval{c_{i} c_{j}^\dagger }[\mathcal{G}_2(t)]_{m j}\bigg]
\end{align}	  
\end{widetext}
where $\langle c_p^\dagger c_q \rangle$ and $\langle c_p c_q^\dagger \rangle$, $p, q$ denoting sites anywhere in the entire system, are the initial correlations of the system which are determined by the initial states of the reservoirs and the wire. Note that the full solution depends on the initial state of the wire. We assume that the wire operators are initially un-correlated i.e. $\langle c_i^\dagger c_j \rangle=0$ and $\langle c_i c_j^\dagger\rangle=\delta_{ij}$ which is the same as choosing the initial state of the wire to be $\ket{0}\bra{0}$. The expressions in Eq.~(\ref{G1W}-\ref{G2WR}) enable the use of a similar algebra as in Ref.~\onlinecite{dhar2006}  to obtain,
\begin{align}
	\langle c_l^\dagger(t)c_m(t)\rangle&=N_{lm}^{SS}+N_{lm}^{BS}(t)~~\text{and}\\
	\expval{c_l(t)c_m(t)}&=M_{lm}^{SS}+M_{lm}^{BS}(t),
\end{align}
where $N_{lm}^{SS}$ and $M_{lm}^{SS}$ are the steady state contributions given by Eq.~\ref{denexp} and Eq.~\ref{cccorr} respectively. $N_{lm}^{BS}(t)$ and $M_{lm}^{BS}(t)$  are the bound state contributions to these correlators which are defined by Eq.~\ref{Nlm} and Eq.~\ref{Mlm} respectively.

\end{document}